\documentclass[preprint,aps,superscriptaddress,nofootinbib,showpacs]{revtex4-2}

\usepackage{amsmath,amsfonts,amssymb}
\usepackage{graphicx}
\usepackage{epsfig}
\usepackage{bm}
\usepackage{wrapfig}
\usepackage[usenames, dvipsnames]{color}
\usepackage{ulem}

\usepackage{color} 

\newcommand{\cM}{{\cal M}}

\newcommand{\psibar}{\mbox{$\overline{\psi}$}}

\newcommand{\tldF}{\mbox{${\tilde F}$}}

\newcommand{\epsi}{\mbox{$\varepsilon$}}

\newcommand{\vp}{\mbox{$\bm{p}$}}
\newcommand{\vq}{\mbox{$\bm{q}$}}
\newcommand{\vbr}{\mbox{$\bm{r}$}}

\newcommand{\vB}{\mbox{$\bm{B}$}}

\newcommand{\valph}{\mbox{$\bm{\alpha}$}}

\newcommand{\vepsi}{\mbox{$\bm{\epsilon}$}}
\newcommand{\vlambda}{\mbox{$\bm{\lambda}$}}

\begin{document}

\title{Anisotropic Coulomb Interaction in Strongly Magnetized Electron Gas 
at Zero Temperature in Relativistic Quantum Framework}

\author{Tomoyuki~Maruyama}
\email{maruyama.tomoyuki@nihon-u.ac.jp}
\affiliation{College of Bioresource Sciences,
Nihon University,
Fujisawa 252-8510, Japan}
\affiliation{ Department of Physics, 
Tokyo Metropolitan university, Hachioji, Tokyo 181-8588, Japan}

\author{A.~Baha~Balantekin}
\email{baha@physics.wisc.edu}
\affiliation{Department of Physics, University of Wisconsin, Madison,
WI 53706, USA}

\author{Myung-Ki Cheoun}
\email{cheoun@ssu.ac.kr}
\affiliation{Department of Physics and Origin of Matter and Evolution of Galaxies Institute, Soongsil University, Seoul, 06978, Republic of Korea}

\author{Toshitaka~Kajino}
\email{kajino.cosmo.only1@gmail.com}
\affiliation{School of Physics, Peng Huanwu Collaborative Center for Research and Education, and
International Research Center for Big-Bang Cosmology and Element Genesis, Beihang University,
Beijing 100191, China}
\affiliation{National Astronomical Observatory of Japan, 2-21-1 Osawa, 
Mitaka, Tokyo 181-8588, Japan}
\affiliation{Faculty of Science,
The University of Tokyo, Hongo 7-3-1, Bunkyo-ku, Tokyo 113-0033, Japan}

\author{Eunseok Hwang}
\affiliation{Department of Physics, Soongsil University, Seoul,
156-743, Korea}



\date{\today}


\begin{abstract}
We present a microscopic derivation of Debye screening in a Landau-quantized electron gas 
subject to a strong magnetic field. 
By calculating the photon self-energy within the RPA framework, we explore how Landau levels influence the screening properties of the electron gas. 
Due to the quantum periodicity of Landau levels, the screening behavior is relatively 
insensitive to variations in the carrier density within each occupation cycle compared to the drastic changes induced by the magnetic field strength. 
Within a single occupation cycle, the Kohn anomaly dominates the low-density regime; at higher densities, 
its characteristic momentum shifts into the asymptotic high-momentum regime, 
rendering its singular contribution less prominent over the active screening domain and allowing 
a smooth orientation-dependent anisotropy to govern the system. 
Furthermore, a subtle competition in directional screening is observed between the parallel and perpendicular directions 
to the field depending on the density domain. 
At lower field strengths, multiple quantum oscillations emerge from the simultaneous participation of various Landau levels, marking a qualitative transition toward the isotropic three-dimensional limit.
Our findings reveal anisotropic screening effects and provide insights into the behavior of charged impurities in such systems.
\end{abstract}

\maketitle


\newpage


\section{Introduction}

The phenomenon of electric charge screening is fundamental to understanding the behavior of charged particles in various physical environments. In unmagnetized systems, the classic Debye screening effect arises from the collective response of charge carriers to perturbations \cite{Escande2014}, where the characteristic screening scale is mostly determined by the Debye mass, which is directly related to the electronic compressibility \cite{DbMsRPA}.

In the presence of a magnetic field, however, the collective response of charged particles is significantly modified. Early kinetic studies demonstrated that magnetic fields qualitatively alter plasma collective modes and generate anisotropic electrostatic responses \cite{Harris1959,Harris1965}. In weakly magnetized systems, screening remains well described by the isotropic Debye mechanism \cite{DbMsRPA}, where the dielectric response is determined primarily by the carrier density and temperature. As the magnetic field strength increases, however, the transverse motion of charged particles becomes quantized into discrete Landau levels \cite{LLevel}, leading to a substantial modification of the density of states near the Fermi surface. This quantization introduces new characteristic momentum scales, alters the polarization function, and generates strongly anisotropic dielectric properties.

Another crucial consequence of a sharp Fermi surface is the emergence of spatial quantum singular structures, namely the Kohn anomaly \cite{Kohn1959} and Friedel oscillations \cite{Friedel1958}. 
These originate from the non-analytic behavior of the static density response at a momentum transfer connecting opposite points of the Fermi surface. In an ordinary three-dimensional electron gas, this anomaly appears at a momentum transfer twice the Fermi momentum and is closely related to Friedel oscillations in real space. 
In a strong magnetic field, where transverse motion is quantized into Landau levels while longitudinal motion along the field remains continuous, this quantum effect becomes highly anisotropic \cite{Horing69, Potekhin99}. 
As a result, the Kohn anomaly acquires an effectively one-dimensional character, 
appearing when the momentum transfer parallel to the magnetic field satisfies the one-dimensional nesting condition 
 connecting Fermi points within each occupied Landau level. 
In addition to the finite-momentum Kohn anomaly, the system exhibits threshold singularities associated with the enhanced one-dimensional density of states near the Landau-level bottoms. 
This singular finite-momentum response plays a central role in the anisotropic screening behavior discussed in the present work.

Such Landau quantization effects have been extensively discussed in the context of magnetized plasmas \cite{GC1968,Horing69}, quantum oscillatory phenomena in metals \cite{dHvA1930, SHeff30, LK56}, and strongly magnetized neutron-star crusts \cite{Potekhin99, Sharma2010,Chamel2008,Baiko2009,EMS2013}, where they influence collective excitations, transport coefficients, lattice stability, and impurity interactions. Consequently, understanding the microscopic screening mechanism in a Landau-quantized electron gas remains a vital problem in both plasma physics and astrophysics.

Previous studies have investigated the influence of strong magnetic fields on plasma properties, particularly focusing on the quantization 
of electron motion into Landau levels \cite{LLevel} and its consequences for collective excitations and damping mechanisms \cite{Sharma2010}. 
It has been shown that under extreme magnetic fields, such as those found in magnetars, electrons occupy only the lowest Landau level, leading to highly anisotropic screening effects. 
The screened potential between static charges exhibits oscillatory behavior parallel to the magnetic field, while the damping of lattice phonons becomes strongly direction-dependent. 
Most of these analyses, however, rely on specific astrophysical conditions or employ simplified semiclassical descriptions, leaving open questions regarding a more general microscopic formulation applicable to a wide range of magnetized electron systems.

More recently, the screening properties of relativistic magnetized plasmas have been investigated 
in astrophysical environments. 
Magnetic-field modifications of the Debye screening length and their consequences 
for weak interaction rates, electron capture processes, and nucleosynthesis 
in Landau-quantized electron-positron plasmas have been evaluated \cite{Famiano2020,Luo2020,Luo2022}. 
Furthermore, the microscopic photon polarization tensor and photon self-energy in strong magnetic fields 
have been systematically evaluated beyond the lowest Landau level regime \cite{Karbstein2013,Hattori2013,Hattori2022,Fukushima2025}. 
In particular, the full formulation of the density-density response function incorporating arbitrary Landau levels has been established in Ref.~\cite{Fukushima2025}, which in the static limit directly corresponds to the Coulomb correlation function evaluated in the present framework.

Such anisotropic screening may affect impurity scattering, transport coefficients, lattice stability, and pycnonuclear reaction rates in strongly magnetized neutron-star crusts \cite{Chamel2008,Baiko2009}. While our primary focus is on the electrostatic Coulomb interaction, we acknowledge the complexity introduced by magnetic interactions. The inclusion of magnetic effects in the screening process presents significant numerical challenges, which we defer to future studies.

The theoretical formalism based on the Green's function method is presented in Sec.~II. Numerical results and discussions are given in Sec.~III, and our conclusions are summarized in Sec.~IV.

\section{Formalism }

\subsection{Electron Green Function in Strong Magnetic Field}

We assume a uniform constant magnetic field along the $z$-direction,
$\vB = (0,0,B)$, and take the electro-magnetic vector potential $A^{\mu}$ to be
$A = (0, 0, x B, 0)$ at the position $\vbr \equiv (x, y, z)$ .

The relativistic electron wave function $\psi$ in the magnetic field is obtained
from the following Dirac equation:
\begin{equation}
\left[ \gamma_\mu \cdot (i \partial^\mu + e A^\mu) - m_e
\right]
\psi (x) = 0 ,
\label{DirEq}
\end{equation}
where $m_e$ is the electron mass, and $e$ is the elementary charge.
Here, we write the wave function as
\begin{equation}
 \psi(X) = \left( \begin{array}{c}
\lambda_1 f_{n_L - 1} \left( \sqrt{eB} x + p_y / \sqrt{eB} \right)  \\
 \lambda_2 ~ f_{n_L} \left( \sqrt{eB} x + p_y / \sqrt{eB} \right) ~  \\
\lambda_3 f_{n_L - 1} \left( \sqrt{eB} x + p_y / \sqrt{eB} \right) 
  \\ \lambda_4 ~ f_{n_L } \left( \sqrt{eB} x + p_y / \sqrt{eB} \right) ~
\end{array}\right)  e^{i (p_y y + p_z z -  p_0 t) } .
\end{equation}
and rewrite Eq.~(\ref{DirEq}) as  the following characteristic equation:
\begin{eqnarray}
&& \left[ p_0 \gamma_0 -  \sqrt{2 eB n_L} \gamma^2 - p_z \gamma^3
- m_e \right] \vlambda
= 0
\label{ChrEq}
\end{eqnarray}
with $\vlambda = ^t (\lambda_1, \lambda_2, \lambda_3, \lambda_4)$,
where $n_L $ is the Landau level number, and $f_{-1} = 0$
By solving Eq.~(\ref{ChrEq}), we then obtain the energy eigenvalues as
\begin{equation}
p_0 = \pm E(n_L, p_z) = \pm \sqrt{ 2 eB n_L + p_z^2 + m_e^2 }.
\label{SgEn}
\end{equation}
Therefore the transverse energy scale associated with Landau quantization is set by 
$\sqrt{eB}$.
The electron Green function in the three dimensional real space is written as
\begin{equation}
 S (\vbr ,\vbr^\prime, E) = \sum_n \psi_n(\vbr) \psibar_n (\vbr^\prime)
\left[ \frac{\theta(E_n - E_F)}{E - E_n + i \delta} 
+ \frac{\theta(E_F - E_n)}{E - E_n - i \delta} \right]
\end{equation}
where $n$ indicates all quantum number describing electron states,
$E$ is the transfer energy, $\delta$ is the minimal positive constant,
and $E_F$ is the Fermi energy.
We should note that the latter part includes negative energy states
as well as occupied states with  positive energies.

After a Fourier transformation with respect to $y$ and $z$ direction,
we can rewrite the Green Function as
\begin{equation}
 S (\xi,\xi^\prime, p_z, p_0) = S_F (\xi,\xi^\prime, p_z, p_0) 
+ S_D (\xi,\xi^\prime, p_z, p_0) 
\label{ElGr}
\end{equation}
with $\xi = x + p_y/eB$,
where the former and latter parts are the Feynman and density-dependent
parts, respectively.
These parts are given by
\begin{eqnarray}
S_F  (\xi,\xi^\prime, p_z, p_0) &=&  \sum_{n_L=0} \tldF_{n_L} (\xi)  
\frac{ \rho_M (p_0, p_z, n_L) }{ \left[ p_0^2 - E^2  + i \delta \right] } \tldF_{n_L} (\xi^\prime) , 
\label{ElGrF}
\\
S_D  (\xi,\xi^\prime, p_z, p_0) &=& \sum_{n_L=0} \frac{i \pi }{E} 
\tldF_{n_L} (\xi) \rho_M (E, p_z, n_L) \tldF_{n_L} (\xi^\prime) 
n_e (E) \delta \left[ p_0 - E (n_L, p_z)  \right] 
\label{ElGrD}
\end{eqnarray}
with $n_e(E) = \Theta (E_F - E)$, and
\begin{eqnarray}
&& \tldF_{n_L} =  {\rm diag} \left( f_{n_L-1}, f_{n_L}, 
f_{n_L-1}, f_{n_L} \right)
= f_{n_L-1}  \frac{1 + \Sigma_z }{2} +  f_{n_L}  \frac{1 - \Sigma_z }{2}  .
%
\\
&& \rho_M (p_0, p_z, n) = p_0 \gamma_0 - \sqrt{2 eB n_L} \gamma^2  - p_z \gamma^3 + m_e 
 \end{eqnarray}
where $\Sigma_z = diag(1,-1,1,-1)$, and
 $n$ and $s$ are a Landau number and a spin of an electron, respectively.


\subsection{Photon Propagator in Medium}

We define the propagator of the photon with momentum $q$ as $D_{\mu \nu} (q)$
in electron gas,
Using the randam phase approximation (RPA) this photon propagtor with momentum $q$
satisfies the following equation.
\begin{equation}
D_{\mu \nu} (X, X^\prime)  = 
\Delta_{\mu \nu} (X, X^\prime)  + \int d^4 Z d^4 Z^\prime  
\Delta_{\mu \alpha} (X, Z)  G^{\alpha \beta} (Z, Z^\prime) 
 D_{\beta \nu} (Z^\prime, X^\prime) ,
\label{PhPrR}
\end{equation}
where $\Delta$ is the photon propagator in vacuum, and
and $G$ is the correlation function of electrons.
In this work we calculate the correlation function $G$ in the RPA, which includes  the one-loop electron contribution, is given by
\begin{equation}
G_{\mu \nu} = -i {\rm Tr} \left\{ \gamma_\mu S(X, X^\prime) 
\gamma_\nu S(X^\prime, X) \right\} .
\end{equation}

Using the Fourier transformation of Eq.~(\ref{PhPrR}),  we obtain the photon propagator
with momentum $q$ in the momentum space as
\begin{equation}
D_{\mu \nu} (q)  = 
\Delta_{\mu \nu} (q)  + \Delta_{\mu \alpha} (q)  G^{\alpha \beta} (q)  D_{\beta \nu} (q) 
\label{PhPrEq}
\end{equation}
with
\begin{equation}
\Delta (q) = \left(- g_{\mu \nu} + \frac{q_\mu q_\nu}{q^2} \right) \frac{1}{ q^2 + i \delta } 
\end{equation}

The system has a rotational symmetry around the $z$-axis, and we can set the photon momentum
to be $q= (q_0; \vq) = (q_0; 0, q_T, q_z)$ without loss of generality.
Then, the correlation function in the momentum space is written as
\begin{equation}
G_{\mu \nu} (q) =
\frac{-i (eB)^2 }{(2 \pi)^2} \sum_{n_i, n_f} \int dp_z dp_0 d x d x^\prime
{\rm Tr} \left\{ \gamma_\mu  S(\xi, \xi^\prime, p+q, n_f) \gamma_\nu  S(\xi^\prime, \xi, p, n_i) \right\} .
\label{GcorP}
\end{equation}

Here, we choose the Lorentz gauge, where $q_\mu \Delta^{\mu \nu}= 0$, and the current conservation
leads to $q_\mu G^{\mu \nu} = 0$, so that $q_\mu D^{\mu \nu}= 0$ must be satified.
For convenience, then,  define the polarization vectors $\epsilon$, which must satisfy
$\epsilon \cdot q = 0$, as follows.
\begin{equation}
\epsilon (L) = \frac{1}{\sqrt{q^2}} \left(|\vq|; \frac{q_0 \vq}{|\vq|} \right) ,\quad
\epsilon(X) = (0; 1, 0, 0 ), \qquad \epsilon (T)= \frac{1}{|\vq|} (0; 0, q_z, -q_T) ,
\label{PolV}
\end{equation}
The photon propagator in vacuum can be rewritten as
\begin{equation}
\Delta^{\mu \nu} (q) =  \frac{1}{q^2 + i \delta}
\sum_{\alpha = {L, X, T} } \epsilon^\mu (\alpha) \epsilon^\nu (\alpha) .
\end{equation}
Then, we could describe the photon propagator $D$ and the correlation function $G$
in terms of the above polarization vectors as
\begin{equation}
D_{\alpha, \beta} = \epsilon^\mu (\alpha) D_{\mu \nu} \epsilon^\nu (\beta) ,
\quad
G_{\alpha, \beta} = \epsilon^\mu (\alpha) G_{\mu \nu} \epsilon^\nu (\beta) .
\end{equation}
Without a magnetic field,  due to the spherical symmetry of the system, 
the correlation function is diagnalized for the above polarization vectors.
and degenerate with respect to the transverse vectors, $\epsilon(X)$ and $\epsilon(T)$. 
With a magnetic field, however, the spherical symmetry is broken,  
the diagonality is unclear, and the degeneracy is resolved.

Although the magnetic fields relevant to the primary scope of this work are focused
 on the subcritical regime ($B \le 10^{12}$~G),  which is well below the QED critical field 
 $B_c = m_e^2 / e \approx 4.414 \times 10^{13}$~G, 
it is worth noting that within this field domain, vacuum polarization effects \cite{Tsai1974, Dittrich2000, Karbstein2013} 
are higher-order corrections that remain negligible 
compared to the dominant density-dependent medium contribution responsible for Debye screening. 
Therefore, the QED vacuum contribution is omitted throughout the present work to focus purely on the medium-induced dielectric response.
%

\subsection{Screening Potential}

In this subsection, we consider the screening potential between two baryons.

The lowest interaction with respect to the fine structure constant is described by
the one-photon interaction in electron gas.
In addition we assume that the energy transfer in the interaction is negligible because
baryons do not  do not move much in this system.   

When $q_0=0$, the polarization vectors are written as  $\epsilon_L = (1;0,0,0)$ for the L mode, 
and as $\vepsi_X = ( {\hat z} \times {\hat q} )/ \left| {\hat z} \times {\hat q} \right|$ 
and $\vepsi_T =  ( {\hat q} \times \vepsi_X )/ \left| {\hat q} \times \vepsi_X  \right|$ 
for the X and T modes ,
where $\epsilon_\alpha = (0; \vepsi_\alpha)$ for $\alpha = T, X$.
  
In static vacuum system the electro-magnetic interaction energy  between two baryons
is written in the momentum space as
\begin{equation}
V^{(0)}_{e,m.}  =  e^2 \int \frac{d^3 \vp}{ (2 \pi)^3 } \frac{d^3 \vp^\prime }{ (2 \pi)^3 } 
\bar{\hat \psi}_B (\vp^\prime) \gamma^\mu {\hat \psi}_B (\vp) 
\left[ - \Delta_{\mu \nu} (0; \vp^\prime - \vp ) \right]
\bar{\hat \psi}_B (\vp)  \gamma^\nu  {\hat \psi}_B (\vp^\prime)
\end{equation}
with ${\hat \psi}_B$ being  the baryon field.
In medium, we replace the photon propagator in medium, $D$, from $\Delta$ and 
write
\begin{equation}
V_{e,m.}  =  e^2 \int \frac{d^3 \vp}{ (2 \pi)^3 } \frac{d^3 \vp^\prime }{ (2 \pi)^3 } 
\bar{\hat \psi}_B (\vp^\prime) \gamma^\mu {\hat \psi}_B (\vp)
\left[ - D_{\mu \nu} (0; \vp^\prime - \vp ) \right]
\bar{\hat \psi}_B (\vp)  \gamma^\nu  {\hat \psi}_B (\vp^\prime) .
\end{equation}
Thus, $- D(q)$ indicates the two body potential in the momentum space.

In the static limit ($q_0 = 0$), 
all non-diagonal components of the correlation function vanish, $G_{\alpha\beta}(0, \bm{q}) = 0$ ($\alpha \neq \beta$).
Consequently, the L, X and T modes completely decouple from one another. 
The photon propagator in each independent channel can thus be conveniently diagonalized as
\begin{equation}
-D_{\alpha}^{-1}(\bm{q}) = |\bm{q}|^2 + e^2 G_{\alpha}(\bm{q}),
\end{equation}
where $D_\alpha \equiv D_{\alpha, \alpha}$ and $G_\alpha \equiv G_{\alpha, \alpha}$ denote the diagonal components in the respective mode space ($\alpha = 0, 1, 2, 3$), with $G_0(\bm{q}) \equiv G_{00}(0, \bm{q})$ representing the Coulomb channel relevant for static screening.

Therefore, the electro-magnetic interaction can be rewritten as
\begin{eqnarray}
V_{e,m.}  &=&  e^2 \int \frac{d^3 \vp}{ (2 \pi)^3 } \frac{d^3 \vq}{ (2 \pi)^3 } 
{\hat \psi}^\dagger_B (\vp + \vq) {\hat \psi}_B (\vp)
\left[ - D_L (0; \vq) \right]
{\hat \psi}^\dagger_B (\vp)  \gamma^0  {\hat \psi}_B (\vp + \vq) 
%
\nonumber \\ &+&  e^2 \int \frac{d^3 \vp}{ (2 \pi)^3 } \frac{d^3 \vq}{ (2 \pi)^3 } 
{\hat \psi}^\dagger_B \left( \vepsi_T \cdot \valph \right) {\hat \psi}_B \left[ - D_{T} (0; \vq) \right]
{\hat \psi}^\dagger_B \left( \vepsi_T \cdot \valph \right) {\hat \psi}_B  
%
\nonumber \\ &+&  e^2 \int \frac{d^3 \vp}{ (2 \pi)^3 } \frac{d^3 \vq}{ (2 \pi)^3 } 
{\hat \psi}^\dagger_B \left( \vepsi_X \cdot \valph \right) {\hat \psi}_B \left[ - D_{X} (0; \vq) \right]
{\hat \psi}^\dagger_B \left( \vepsi_X \cdot \valph \right) {\hat \psi}_B  .
\label{ModeVem}
\end{eqnarray}
In the above equation, the first term shows the Coulomb interaction,
and the second and third terms indicate the magnetic interaction.
Though  these two terms are the same without an external magnetic field, 
they give different contributions to the interaction energy in a strong magnetic field.  The  $\vepsi_T$ is in the plane composing the $\vq$ and $\vB$, and the $\vepsi_X$ is  perpendicular to this plane.  Thus, we call the $L$, $T$ and $X$ modes the coulomb mode, the in-plane magnetic mode  and the out-of-plane magnetic mode, respectively, and rewrite $\vepsi_T \equiv \vepsi_{in}$ and $\vepsi_X \equiv \vepsi_{out}$.  
In the present work we restrict the numerical analysis to the Coulomb mode $G_C$, while the magnetic modes are retained only to define the complete tensor decomposition.

In the present work, we focus on the static density-density response and evaluate the Coulomb screening 
within the longitudinal electric channel of the RPA.  Possible mixing with transverse electromagnetic modes 
is not included in the present calculation and requires a separate analysis of the full polarization tensor.


\subsection{Correlation Function}

In this subsection we give the expressions of the correlation function in the RPA.

As mentioned before, we can write the photon momentum
to $q= (q_0; \vq) = (q_0; 0, q_T, q_z)$ without losing the generality,
and use the photon polarization function. 

As written before, the electron propagator is separetaed into two parts as Eq.~(\ref{ElGr}).
Along this line we also separate the correlation funciton into two parts as
\begin{equation}
G^{\mu \nu} (q) = G_F^{\mu \nu} (q) + G_D^{\mu \nu} (q) ,
\label{CorSp}
\end{equation}
where the first term is called the Fyenman part which is defined as
\begin{equation}
G_F^{\mu \nu}  (q) = \frac{-i (eB)^2 }{ ( 2 \pi)^2 } \sum_{n_i. n_f} \int dp_0 dp_z d x d x^\prime
{\rm Tr}  \left\{ \gamma_\mu S_F (\xi, \xi^\prime,  p+q, n_f) 
\gamma_\nu S_F (\xi^\prime, \xi, p, n_i) \right\} ,
\end{equation}
and the second term of Eq.~(\ref{CorSp}) is called the density-dependent part.
which is written as
\begin{equation}
G_D^{\mu \nu} (q) = G_D^{(1) \mu \nu} (q) +  G_D^{(2) \mu \nu} (q)  + G_D^{(3) \mu \nu} (q)
\end{equation}
with
\begin{eqnarray}
G_D^{(1) \mu \nu}  (q)  &=& \frac{-i (eB)^2 }{ ( 2 \pi)^2 } \sum_{n_i. n_f} \int dp_0 dp_z d x d x^\prime
{\rm Tr}  \left\{ \gamma_\mu S_F (\xi, \xi^\prime, p+q, n_f) 
\gamma_\nu S_D (\xi^\prime, \xi, p, n_i) \right\} ,
\label{Gcor1}
\\
G_D^{(2) \mu \nu}  (q)  &=& \frac{-i (eB)^2}{ ( 2 \pi)^2 } \sum_{n_i. n_f} \int dp_0 dp_z d x d x^\prime 
{\rm Tr}  \left\{ \gamma_\mu S_D (\xi, \xi^\prime, p, n_i) 
\gamma_\nu S_F (\xi^\prime, \xi, p - q, n_f) \right\} ,
\label{Gcor2}
\\
G_D^{(3) \mu \nu}  (q)  &=& \frac{-i (eB)^2 }{ ( 2 \pi)^2 } \sum_{n_i. n_f} \int dp_0 dp_z d x d x^\prime
{\rm Tr}  \left\{ \gamma_\mu S_D (\xi, \xi^\prime, p+q, n_f) 
\gamma_\nu S_D (\xi^\prime, \xi, p, n_i) \right\} .
\label{Gcor3}
\end{eqnarray}

Substituting Eqs.~(\ref{ElGrF}) and (\ref{ElGrD}) into Eq.~(\ref{Gcor1}), we obtain
\begin{eqnarray}
G_{D \mu \nu}^{(1)}  (q) &=& \frac{- 4i (eB)^2 }{ ( 2 \pi)^2 } \sum_{n_i. n_f} \int dp_0 dp_z 
\frac{i \pi }{E_i} n _e (E_i)  
\frac{ N_{\mu \nu}  (p+q, n_f; p, n_i) \delta \left[ p_0 - E_i (n_i, p_z)  \right] }{(p_0 + q_0)^2 - (p_z + q_z)^2 - 2eB n_f - m_e^2 + i\delta}  
\nonumber \\ &=&
\frac{ (eB) }{ \pi } \sum_{n_i. n_f} \int \frac{ dp_z  }{E_i} n_e (E_i)  
\frac{ N_{\mu \nu} (p+q, n_f; p, n_i)  }{(E_i + q_0)^2 - (p_z + q_z)^2 - 2eB n_f - m_e^2 + i\delta } ,
\\
G_{D \mu \nu}^{(2)}  (q)  &=&
\frac{ eB }{ \pi } \sum_{n_i. n_f} \int \frac{ dp_z  }{E_i} n_e (E_i)  
\frac{ N_{\mu \nu}  (p, n_i; p-q, n_f)  }{(E_i - q_0)^2 - (p_z - q_z)^2 - 2eB n_f - m_e^2 + i\delta }  ,
%
\\
G_{D \mu \nu}^{(3)}  (q)  &=&
 i eB \sum_{n_i. n_f} \int \frac{ dp_z  }{E_i E_f} n_e (E_i)  n_e (E_f)
N_{\mu \nu} (p+q, n_f; p, n_i)   \delta \left[ q_0 + E_i - E_f \right]
\end{eqnarray}
with $E_f = \sqrt{2 eB n_f + (p_z + q_z)^2 + m_e^2}$. 
with $E_i = \sqrt{2 eB n_i + p_z^2 + m_e^2}$, and
\begin{equation}
N_{\mu \nu} (p_f, n_f ; p_i. n_i )=  \frac{1}{4} {\rm Tr} \left\{
 \rho_M (p + q, n_f) J_\mu (n_f, n_i, q_T) \rho_M (p, n_i)  
{\bar J}_\mu  (n_i, n_f, -q_T)  \right\} ,
\end{equation}
where $\rho_M = (p_0 \gamma^0 - p_z \gamma^3 - \sqrt{2eB n} \gamma^2 - m_e)$ and 
\begin{eqnarray}
J_{\mu} (n_f, n_i, q_T) &=& \int d x \tldF(n_f, s_f, x + q_T/2 \gamma_\mu  \tldF(n_i, s_i, x - q_T/2) ,
\label{Jcur}
\\
{\bar J}_{\mu} (n_i, n_f, -q_T) &=& \int d x  \tldF(n_i, s_i, x - q_T/2) \gamma_\mu  \tldF(n_f,s_f,x + q_T/2) .
\label{Jctl}
\end{eqnarray}
%

Here, we define the following overlap function between two one-dimensional harmonic oscillator wave-function:
\begin{eqnarray}
 \cM(n_1,n_2) 
 &=& \int d x f_{n_1}\left( x + \frac{q_T}{2} \right) 
f_{n_2} \left( x - \frac{q_T}{2} \right) 
%
\nonumber \\
 &=& \sqrt{ \frac{n_2!}{n_1! } }
\left( \frac{q_T}{\sqrt{2} } \right)^{n_1-n_2}  
e^{-\frac{q_T^2}{4} } L^{n_1-n_2}_{n_2} \left( \frac{q_T^2}{2} \right)
\quad~ \left( n_1 \le n_2 \right)
\nonumber \\ 
&=&
\sqrt{ \frac{n_1!}{n_2! } }
\left(- \frac{q_T}{\sqrt{2} } \right)^{n_2-n_1}  
e^{-\frac{q_T^2}{4} } L^{n_2-n_1}_{n_1} \left( \frac{q_T^2}{2} \right) 
~~ \left( n_1 \ge n_2 \right)  .
\label{OLHO}
\end{eqnarray}

Substitute Eq.~(\ref{OLHO}) into Eqs.~(\ref{Jcur}) and (\ref{Jctl}), we can obtain
\begin{eqnarray}
J_{\mu} 
&=&  \left[ \cM_1 \frac{1 + \Sigma_z}{2} 
 + \cM_2 \frac{1 - \Sigma_z}{2} \right] \gamma_\mu \quad (\mu = 0, 3)
\\
&=&
 \left[ \cM_3 \frac{1 + \Sigma_z}{2} 
 + \cM_4 \frac{1 - \Sigma_z}{2} \right]  \gamma_\mu \quad (\mu = 1, 2) ,
\end{eqnarray}
and
\begin{eqnarray}
{\bar J}_{\mu} &=&
 \left[ \cM_1 \frac{1 + \Sigma_z}{2} 
 + \cM_2 \frac{1 - \Sigma_z}{2} \right]  \gamma_\mu \quad (\mu = 0, 3)
\label{PolV1}
\\ &=&
 \left[ \cM_4 \frac{1 + \Sigma_z}{2} 
 + \cM_3 \frac{1 - \Sigma_z}{2} \right] \gamma_\mu  \quad (\mu = 1, 2)  .
\label{PolV2}
\end{eqnarray}
with
\begin{eqnarray}
\cM_1= M \left( n_i, n_f \right) , \quad~~~
&&
\cM_2 = M \left( n_i - 1, n_f - 1 \right) ,
\nonumber  \\
\cM_3 = M \left( n_i-1, n_f \right) ,
&&
\cM_4 = M \left( n_i, n_f - 1 \right) .
\end{eqnarray}
%


Next, we show the detailed expressions of $N_{\mu \nu}$ in the following.
As seeing Eqs.~(\ref{PolV}), we should distinguish $\mu = 0, 3$ and
$\mu = 1, 2$; the formar and latter parts are refered to the parallet (P) part sideward (S) 
parts, respectively.
Then, we write $N_P^{\mu \nu}$ when $\mu = 0$ or $=3$, and $\nu = 0$ or $=3$,
 $N_S^{\mu \nu}$ when $\mu = 1$ or $=2$, and $\nu = 1$ or $=2$,
 and $N_M^{\mu \nu}$ otherwise.
In addition, we define $g(P) = {\rm diag}(1, 0, 0, -1)$, $g(S) = {\rm diag}(0, -1, - 1, 0)$ and
$p_{aP} = (E_{ai} ; 0, 0,  p_{az} )$ .
The detaile expressions are given by
\begin{eqnarray}
 N_P^{\mu \nu} (p_f, n_f ; p_i, n_i) 
 &=& \frac{\cM_1^2 + \cM_2^2}{2}
 \left\{ g^{\mu \nu} (P) \left( m_e^2 - p_{f P} \cdot p_{i P}  \right)
+ \left( p_{i P}^{\mu} p_{f \nu} +  p_{f P}^{\mu} p_{i P}^{\nu}  \right)
\right\}
 \nonumber \\ && 
+ 2 eB \cM_1 \cM_2 \sqrt{n_i n_f}  g^{\mu \nu} (P)
\label{NPP}
\\
N_{S}^{\mu \nu} (p_f, n_f ; p_i, n_i) &=&  
 \frac{\cM_3^2 + \cM_4^2}{2}  g^{\mu \nu}(S) \left( m_e^2 -  p_{iP} \cdot p_{fP}  \right)
+ 2 \cM_3 \cM_4  \sqrt{n_i n_f}  \left(  \delta^\mu_2  \delta^\nu_2  -  \delta^\mu_1 \delta^\nu_1 \right)
\nonumber \\ && \qquad\qquad
+ i \left( \frac{\cM_3^2 - \cM_4^2}{2} \right)
 \left(  \delta^\mu_1  \delta^\nu_2  -  \delta^\mu_2 \delta^\nu_1 \right)
(p_{fP} \cdot p_{iP} - m_e^2 ). ~~~
\label{NSS}
\\
N_{M}^{\mu \nu} (p_f, n_f ; p_i, n_i) &=&  
\frac{\cM_1 \cM_4 + \cM_2 \cM_3}{2}   \sqrt{2 eB n_f}  
 \left[ p_{iP}^\mu  \delta^\nu_2 +\delta^\mu_2 p_{iP}^\nu \right] 
 \nonumber \\ &&
+ \frac{\cM_2 \cM_4 + \cM_1 \cM_3}{2}   \sqrt{2 eB n_i} 
 \left[ p_{fP}^\mu  \delta^\nu_2 +\delta^\mu_2 p_{fP}^\nu \right] 
\nonumber \\ &&
+ \frac{ \cM_1 \cM_4 - \cM_2 \cM_3}{2} (i)  \sqrt{2 eB n_f}
\left[  p_{iP}^\mu \delta^\nu_1 -  \delta^\mu_1 p_{iP}^\nu  \right]
\nonumber \\ &&
+ \frac{ \cM_2 \cM_4 - \cM_1 \cM_3 }{2} (i)  \sqrt{2 eB n_i}
 \left[ p_{fP}^\mu \delta^\nu_1 -  \delta^\mu_1 p_{fP}^\nu  \right] .
\label{NMX}
\end{eqnarray}
%
%

Here, we use the polarization vector in Eq.~(\ref{PolV}), substitute Eqs.~(\ref{NPP}), (\ref{NSS}) and (\ref{NMX})
into  Eqs.~(\ref{Gcor1}) and (\ref{Gcor2}), and obtain the real part of the correlation functions.

\bigskip

%

In the present work we study the static screening of the Coulomb interaction, 
so that we focus only $\mu = \nu = 0$ and take the energy transfer to be  $q_0=0$.
In this condition the imaginary part of the correlation function  at $T=0$ is zero, 
${\rm Im} G^{00}_D (q_0  = 0, \vq) = 0 $.
When $p_{i0} = p_{f0} = E_i$, then,
\begin{eqnarray}
N^{00} (p_i+q, n_f ; p_i, n_i) &=& 
 \frac{\cM_1^2 + \cM_2^2}{2} \left( m_e^2 + E_i^2 + p_{iz}^2 + q_z p_{iz}  \right) 
+ 2 eB \cM_1 \cM_2 \sqrt{n_i n_f} ,
\\
N^{00} (p_i, n_i ; p_i - q, n_f) &=& 
 \frac{\cM_1^2 + \cM_2^2}{2} \left( m_e^2 + E_i^2 + p_{iz}^2 - q_z p_{iz}  \right) 
+ 2 eB \cM_1 \cM_2 \sqrt{n_i n_f} . \quad
\label{N00D}
\end{eqnarray}
%
%
Therefore, the correlation function for the Coulomb mode is written as
\begin{eqnarray}
&& G^C_{D} (\vq) \equiv \epsilon^\mu G^C_{\mu \nu} (\vq) \epsilon^\nu 
= - G_D^{00} (\vq) 
%
%
%
%
%
\nonumber \\ &=&
 \frac{ eB }{ \pi^2 } \sum_{n_i=0}^{n_F} \sum_{n_f=0}^{\infty}\int_{0}^{k_z} \frac{ dp_{iz}  }{E_i} 
\frac{ n_e (E_i) }{[ eB (n_f -n_i) + q_z^2/2 ]^2 - q_z^2 p_{iz}^2  }
\nonumber \\ &&  \qquad\qquad\times %
\left\{
\frac{\cM_1^2 + \cM_2^2}{2} 
\left[
 (m_e^2 + E_i^2 + p_{iz}^2) \left( 2eB (n_f -n_i) + q_z^2/2 \right) -q_z^2 p^2_{iz} 
\right]
 \right. \nonumber \\ && \left. \qquad\qquad\qquad\qquad
+ 
2 eB \sqrt{n_i n_f} \cM_1 \cM_2 \left( 2eB (n_f -n_i) + q_z^2/2 \right) 
\right\} ,
\end{eqnarray}
where $k_z (n_i) = \sqrt{ E_F^2 - 2 eB n_i - m_e^2  }$, and 
$n_F = \lfloor (E_F^2-m_e^2)/(2eB) \rfloor$ denotes the maximum Landau level index below the Fermi energy, with $\lfloor x \rfloor$ being the floor function (i.e., the greatest integer less than or equal to $x$.

In the above equation, the integrand includes singular points, and 
we evaluate it taking the Cauchy principal value.
When $q_z \neq 0$, defining $p_c \equiv eB (n_f - n_i) / |q_z| + |q_z| /2$, the integration yields 
the Coulomb correlation function:
\begin{eqnarray}
G^C_D (\bm{q}) & = &
\frac{eB}{\pi^2 q_z } \sum_{n_i = 0}^{n_F}  \frac{1}{ E_q }  
\left\{ \left[ \frac{\cM_1^2 + \cM_2^2}{2} \left( eB n_i + m_e^2 \right)   
+ \cM_1 \cM_2 eB n_i \right]
\ln \left| \frac{2 k_z E_q  + q_z E_F }{ 2k_z E_q  - q_z E_F } \right| 
\right\} 
\nonumber  \\ && 
+ \frac{eB}{ \pi^2 q_z} \sum_{n_i=0}^{n_F} \sum_{n_f (\neq n_i)}^{\infty}
\left\{ \frac{\cM_1^2 + \cM_2^2}{2}  \frac{eB (n_f - n_i)}{q_z} \left[
\frac{p_c}{ E_c}  \ln \left| \frac{k_z E_c  + E_F p_c}{ k_z E_c  - E_F p_c } \right| 
- \ln \left| \frac{ E_F + k_z }{ E_F - k_z} \right| \right]
\right. \nonumber \\ && \qquad \left.
+ \left[ \frac{\cM_1^2 + \cM_2^2}{2} \left( eB n_i  + m_e^2 \right)   
+ \cM_1 \cM_2 eB \sqrt{n_i n_f } \right]
\frac{1}{ E_c}  
\ln \left| \frac{k_z E_c  + E_F p_c}{ k_z E_c  - E_F p_c } \right| 
\right\} ,
\label{Gcol}
\end{eqnarray}
where $E_c = \sqrt{p_c^2 + 2 eB n_i + m_e^2}$, $E_q = \sqrt{ q_z^2/4 + 2eB n_i + m_e^2}$, and $n_F$ denotes the maximum Landau level index below the Fermi energy. 
The correlation function $G^C_D(\bm{q})$ diverges logarithmically when the condition $k_z/E_F = |p_c|/E_c$ or $k_z/E_F = q_z/2 E_q$ is satisfied. 
This static density-density response encodes both the discrete Landau-level spectrum and the one-dimensional nesting condition associated with the longitudinal Fermi motion along the magnetic field.

To gain physical insight, we first examine the longitudinal limit ($q_\perp = 0$). In this spatial configuration, we have $\cM_1 = \cM_2 = \delta_{n_i, n_f}$, and Eq.~(\ref{Gcol}) simplifies to
\begin{eqnarray}
{\rm Re} G^C_D (q_z) & = &
\frac{eB}{\pi^2 q_z } \sum_{n_i =0 }^{n_F}  \frac{\left( 2eB n_i + m_e^2 \right)}{ E_q }  
\ln \left| \frac{2 k_z E_q  + q_z E_F }{ 2k_z E_q  - q_z E_F } \right| .
\label{GcolT0}
\end{eqnarray}
Here, the logarithmic divergence occurs at $q_z = 2 k_z(n_i)$, which corresponds to $E_F = E_q$. 
This finite-momentum singularity represents the classic one-dimensional Kohn anomaly \cite{Kohn1959}, resulting from the complete nesting of the one-dimensional Fermi surface along the magnetic field.

Conversely, in the transverse limit ($q_z = 0$), the Coulomb correlation function reduces to
\begin{eqnarray}
G^C_D (q_\perp) & = &
\frac{ eB }{\pi^2} \sum_{n_i = 0}^{n_F}  \frac{E_F}{ k_z(n_i) \left( 2eB n_i + m_e^2 \right) }   
\left\{ \frac{\cM_1^2 + \cM_2^2}{2} \left( eB n_i + m_e^2 \right)  +  \cM_1 \cM_2 eB n_i  \right\}   
\nonumber \\ & &
+ \frac{1}{\pi^2} \sum_{n_i \neq n_f}^{\infty} \frac{1}{ n_f - n_i }
\left\{ \frac{\cM_1^2 + \cM_2^2}{2} \left( k_z(n_i) E_F + \frac{m_e^2}{2} \ln \left| \frac{ E_F + k_z(n_i) }{E_F - k_z(n_i) } \right| \right)    
\right. \nonumber \\ && \qquad\qquad\qquad\qquad\qquad \left.
+ \cM_1 \cM_2 eB \sqrt{n_i n_f }  \ln \left| \frac{ E_F + k_z(n_i) }{E_F - k_z(n_i) } \right| 
\right\} .
\label{Gcolz0}
\end{eqnarray}
Unlike the longitudinal response, the transverse correlation function in Eq.~(\ref{Gcolz0}) exhibits no finite-momentum logarithmic singularity at any $q_\perp$, indicating the absence of the Kohn anomaly in the transverse direction. 
However, a threshold singularity of the form $\sim 1/k_z(n_i)$ appears in the intraband term when $k_z(n_i) \to 0$. 
This feature originates from the divergence of the one-dimensional density of states at the bottom of a Landau level and must be clearly distinguished from the finite-momentum Kohn anomaly at $q_z = 2 k_z(n_i)$.

\subsection{Screening Mass and Permittivity}

For the Coulomb mode, then, we define the screening mass, 
which is so called Debye screening mass, as $M_D^2 \equiv  e^2 G^C (\vq = \bm{0}) $.
The Coulomb correlation function $G^C (q)$ corresponds to the density-density response function of the Landau-quantized electron gas. 
In the long-wavelength limit ($q \rightarrow$ 0), it reduces to the electronic  compressibility, 
$G^C (0) = \partial \rho_e / \partial \mu_e$, 
thereby determining the Debye screening mass through $M_D^2 = e^2 G_C (0)$.

Substituting $q_T = 0$ into Eq.~(\ref{Gcolz0}), we  obtain 
\begin{equation}
M_D^2 = e^2 G^C_D (0; \vq=0)
 =  \frac{ e^3 B }{2\pi^2} \sum_{n_i=0}^{n_F} (2 - \delta_{n_i 0} )  \frac{E_F}{ k_z (n_i) }    
\end{equation}
where we note $\cM_1 = \cM_2 = \delta_{n_i, n_f}$ at $q_T = 0$. 
Here, we write $k_F \equiv k_z (n_F)$ which satsifies the relation as
\begin{equation}
k_z^2 (n_L) = k_F^2 + 2 eB (n_F - n_L).
 \end{equation}
 Then we rewrite the Debye screening mass as
\begin{equation}
M_D^2  =  \frac{ e^3 B }{2 \pi^2} \sum_{n_i=0}^{n_F}  \frac{  (2 - \delta_{n\i 0} ) E_F}{ \sqrt{2 eB (n_F -n_i) + k_F^2} }  .
\end{equation}
This equation indicates that the Debye mass $M_D$ diverges when $k_F = 0$. 
This divergence is a consequence of the ideal $T=0$ density of states at a Landau-level threshold and would be smeared by finite temperature, disorder, or finite Landau-level width. It reflects the singular density of states associated with the opening of a new Landau level.

At the limit of $B=0$, we transfer
 $2 eB \sum_{n_i} \rightarrow \int dp_T p_T$ and get
\begin{equation}
M_D^2 =   \frac{ e^2 }{2 \pi^2} \int_0^{p_F}  dp_T p_T  \frac{E_F}{ \sqrt{p_F^2 - p_T^2} }    
= \frac{e^2}{\pi^2} p_F E_F ,
\label{DBnsB0}
\end{equation}
which is equivalent to the Debye screening mass in the usual RPA in non-magnetic system
\cite{DbMsRPA}. The Coulomb permittivity in the momentum space, $\epsi_C$, is given by
\begin{equation}
- D_{0}^{-1} (0; \vq) = q_s^2 + e^2 G^C(0; \vq) = \epsi_C (\vq) q_s^2
\label{CPerm}
\end{equation}
with $q_s = |\vq|$.
In order to discuss the asymmetry of the permittivity caused by the magnetic field,
we need to isolate the contribution of the Debye mass,
Then, we define the reduced effective function $\epsi_R$  
as follows 
\begin{equation}
- D_{0}^{-1} (0; \vq) = |\vq|^2 + e^2 G^C(0; \vq) = \epsi_R (\vq) \left( q_s^2 + M_D^2 \right) .
\label{RPerm}
\end{equation}
Later, we discuss the asymmetry by using the asymmetric parameter 
for the screening deviation $\delta_{SC}$ defined as
\begin{equation}
\delta_{SC} = \frac{1}{\epsi_R (\vq)} - 1 
= \frac{ M_D^2 - e^2 G^C(0; \vq)  }{ q_s^2 + e^2 G^C(0; \vq)} ,
\label{ScDr}
\end{equation}


\subsection{Correlation Function at $B=0$}

Here, we describe the correlation functions with zero magnetic field, $B=0$. When $B=0$,
the real part of the correlation funcion $G_D$ is written as
\begin{eqnarray}
{\rm Re} G_D
&=& \frac{1}{4 \pi^3}   \int \frac{d^3 p}{E(\vp)} n (E) 
\left\{ \frac{ 2 p^\mu p^\nu + p^\mu  q^\nu +  q^\mu p^\nu  - (p \cdot q) g^{\mu \nu} }{2 p\cdot q + q^2}%
\right. \nonumber \\ &&\left.  \qquad\qquad\qquad\qquad
+ \frac{ 2 p^\mu p^\nu - p^\mu  q^\nu - q^\mu p^\nu  + (p \cdot q) g^{\mu \nu} }{ - 2 p\cdot q + q^2}
\right\} .
\end{eqnarray}
We substitute  $q_0=0$ into the above equation and obtain the Coulomb component as
\begin{eqnarray}
{\rm Re} G_D^C (0; \vq)  &=& - {\rm Re} G_D^{0 0} (0; \vq) 
= \frac{1}{4 \pi^3}   \int \frac{d^3 p}{E_p} n (E_p)   \left\{  
 \frac{2 E_p^2 + \vp \cdot \vq  }{ 2 \vp \cdot \vq + \vq^2 } 
- \frac{ 2 E_p^2 - \vp \cdot \vq }{2 \vp \cdot \vq - \vq^2 } \right\}
%
\nonumber \\ &=& 
\frac{1}{\pi^2} \int_0^{p_F} \frac{d p p^2}{E_p}      
\left\{ 1 + \frac{1}{ q_s p } \left( E_p^2 -  \frac{q_s^2}{4} \right)  
 \ln \left | \frac{ p + q_s /2 } {p  - q_s /2 } \right|  \right\} . \quad
\label{GcorB0}
\end{eqnarray}
When $q_s=0$,
\begin{eqnarray}
&& M_D^2 = e^2 {\rm Re}G_D^C (\vq=0) 
= \frac{e^2}{\pi^2}   \int_0^{p_F}  d p  \left( \frac{p^2}{E_p} + E_p \right)
= \frac{e^2}{\pi^2} p_F E_F ,
\end{eqnarray}
which is the same as  Eq.~(\ref{DBnsB0}). 
This non-magnetic result serves as the reference baseline for identifying the additional anisotropy and quantum oscillations induced by Landau quantization.

\newpage
\section{Results}

\begin{figure}[ht]
\centering
{\includegraphics[angle=270,scale=0.65]{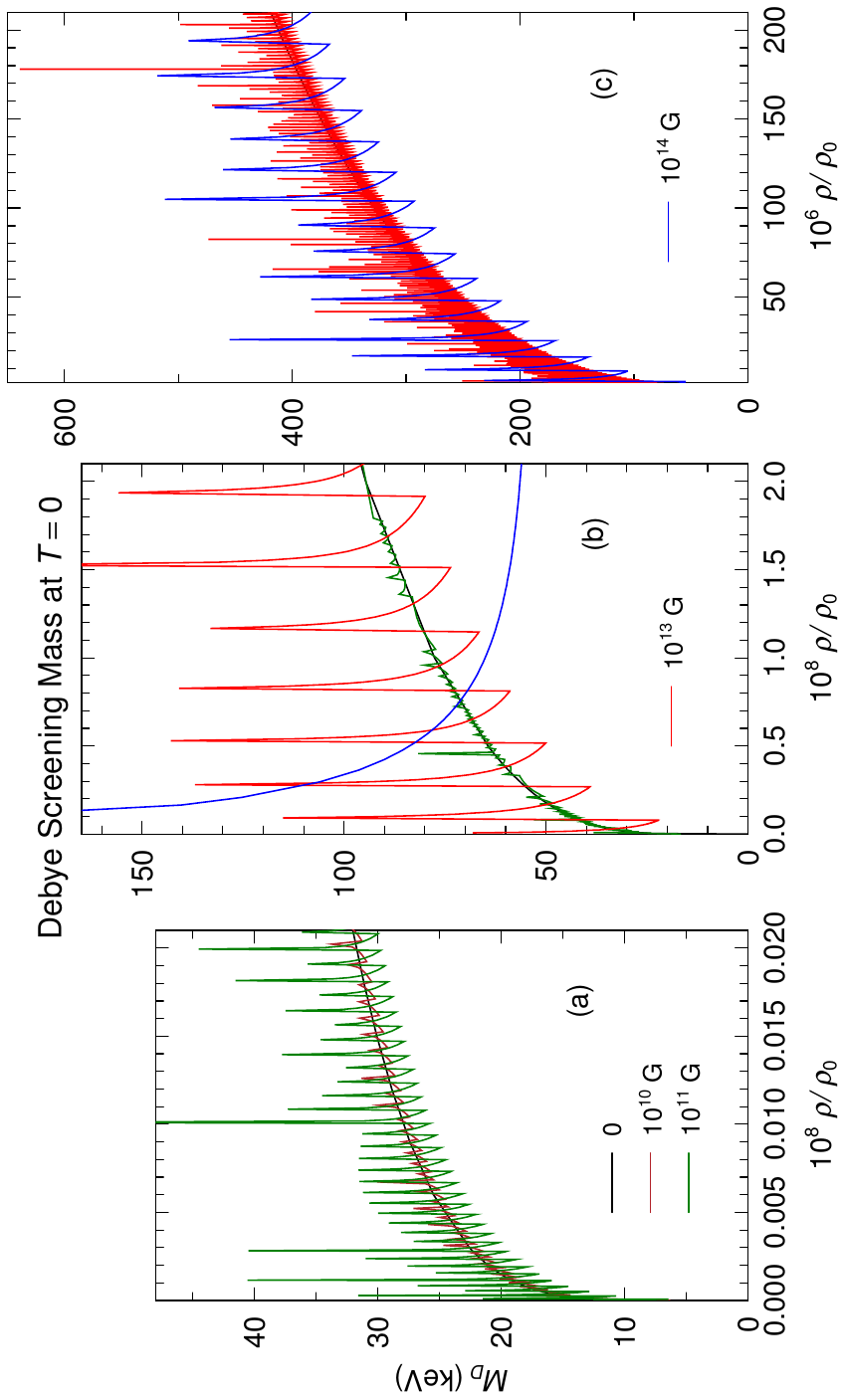} }
\caption{\small
Density dependence of Debye screening mass in magnetized electron gas.
The black, brown, green, red and blue lines represent the results at the magnetic field strengths
of $0$~G, $10^{10}$~G,  $10^{11}$~G,  $10^{13}$~G  and $10^{14}$~G, respectively.
}
\label{ScMas}
\end{figure}

In this section, we present the numerical results at zero temperature. 
In Fig.~\ref{ScMas}, we show the Debye screening mass in a magnetized electron gas, $M_D$, as a function of the electron density normalized by the nuclear saturation density, 
$\rho_0 = 0.17~\mathrm{fm}^{-3}$ ($1.55 \times 10^{14}~\mathrm{g/cm^3}$). 
The left (a), middle (b), and right (c) panels display the same results across different density scales.
While our quantitative analysis is primarily focused on the subcritical field regime ($B \le 10^{12}$~G), 
the case of $B = 10^{14}$~G is also included as an illustrative reference 
to transparently demonstrate the quantum oscillatory structure and 
scaling trends of the pure medium contribution over a broader parameter space.

%

The Debye screening mass exhibits clear quantum oscillations. 
In the ideal $T=0$ limit, the cusps correspond to singular enhancements at Landau-level thresholds, 
{\it i.e.} at the density of $k_F = 0$ $ (E_F^2 = 2 eB n_F + m_e^2)$,
and their intervals are larger as the strength of  magnetic field increases.

\begin{figure}[htb]
\centering
{\includegraphics[scale=0.5]{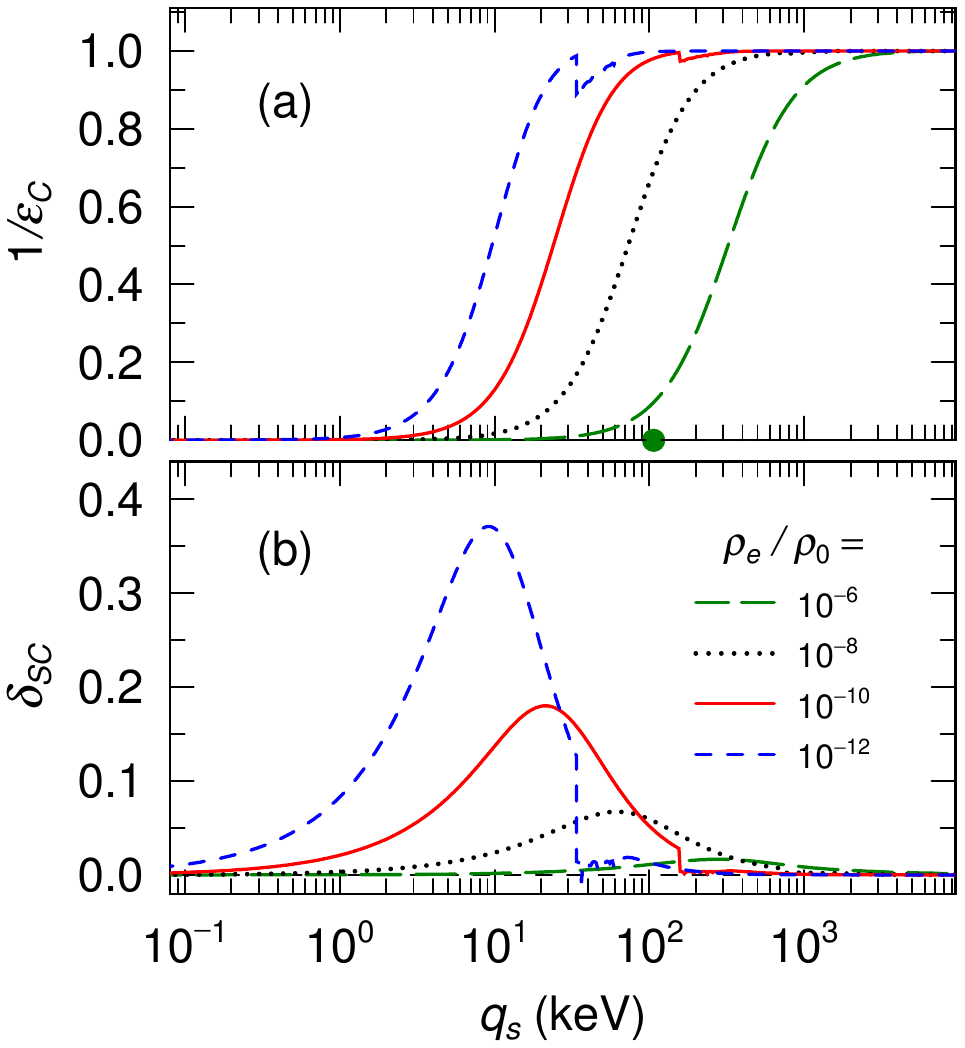} }
\caption{\small
Momentum dependence of the inverse Coulomb permittivity $1/\epsi_C$ (a) and 
the screening deviation $\delta_{SC}$ (b) in a non-magnetized electron gas. 
The blue (dashed), red (solid), black (dotted), and green (long-dashed) lines 
represent the results at electron densities of 
$10^{-12} \rho_0$, $10^{-10} \rho_0$, $10^{-8} \rho_0$, and $10^{-6} \rho_0$, 
respectively.
}
\label{CPmB0}
\end{figure}

In Fig.~\ref{CPmB0}, we show the inverse Coulomb permittivity $1/\epsi_C$ (a) and 
the screening deviation $\delta_{SC}$ (b) in momentum space in the absence of a magnetic field ($B = 0$) at densities of $10^{-12} \rho_0$, $10^{-10}\rho_0$, $10^{-8}\rho_0$, and $10^{-6}\rho_0$. 
The permittivity is defined through $1/\epsi_C = q_s^2/(q_s^2 + e^2 G_C)$ [see Eq.~(\ref{CPerm})]. 
In the limiting cases, one finds $1/\epsi_C \approx 0$ for $q_s \ll M_D$, while $1/\epsi_C \approx 1$ 
for $q_s \gg M_D$. 
Around $q_s \sim M_D$, $1/\epsi_C$ increases rapidly, and correspondingly, 
the screening deviation $\delta_{SC}$ exhibits a peak.
As the density increases, the Debye mass $M_D$ also increases, and the peak position of 
$\delta_{SC}$ shifts toward higher momenta.  
The peak height decreases because the denominator in Eq.~(\ref{ScDr}) becomes larger.
From Eq.~(\ref{DBnsB0}),  we can see that $M_D$ scales with $\rho_e^{1/6}$ in the low-density (non-relativistic) limit where $p_F \ll m_e$, and transitions to a $\rho_e^{1/3}$ scaling in the high-density (ultra-relativistic) limit where $p_F \gg m_e$.

%

\begin{table}[h]
\centering
\begin{tabular}{|c||c|c|c|}
\hline
~$\rho_e$~ &~$10^{10}$~G~ &~$10^{11}$~G~ & ~$10^{12}$~G~ \\ 
\hline 
~$10^{-10} \rho_0 $~ & 208  & ~~20~ & 2 \\ 
\hline 
~~$10^{-9} \rho_0 $~ & 966  & ~~96~ & 9 \\ 
\hline 
~~$10^{-8} \rho_0 $~ & 4486  & ~448~ & 44  \\
\hline 
\end{tabular}
\caption{ 
Maximum occupied Landau-level index $n_F$}
\label{TaLN}
\end{table}

As the correlation function $G^C(0; \vec{q})$ varies largely around the scale of the Fermi momentum $p_F$, the characteristic momentum scale of the electron gas is fundamentally governed by $p_F$. 
The observation that $\delta_{\text{SC}} > 0$ at $B=0$ reflects the inherent momentum dispersion and the $q_s$-dependence of the one-loop RPA polarization function.

We next consider the Coulomb permittivity in magnetic fields.

Table~\ref{TaLN} lists the maximum occupied Landau-level index $n_F$ for several electron densities and magnetic field strengths. 
As shown in Table~\ref{TaLN}, at a fixed electron density, $n_F$ is strongly suppressed as the magnetic field strength increases, reflecting the widening separation between Landau levels.
Conversely, at a fixed magnetic field, higher electron densities allow more Landau levels to be occupied. 
For example, at $\rho_e/\rho_0=10^{-10}$, $n_F$ decreases drastically from $208$ at $B=10^{10}\,\mathrm{G}$ to $2$ at $B=10^{12}\,\mathrm{G}$. 
These results illustrate the transition from a regime with many occupied Landau levels toward one dominated by only a few levels, providing a clear basis for selecting the representative occupation regimes examined in the following figures.

\begin{figure}[ht]
\centering
\includegraphics[scale=0.5]{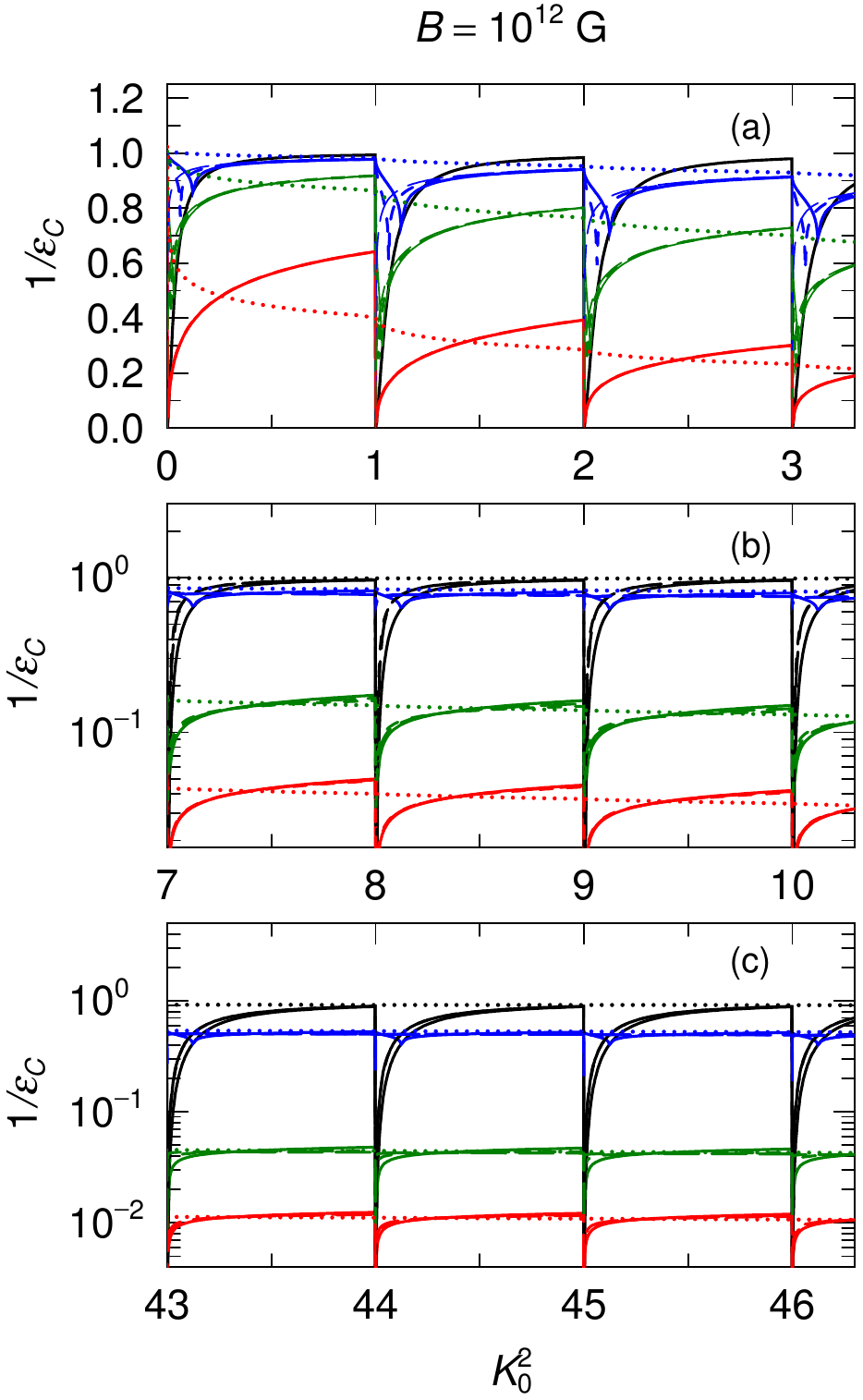}
\caption{\small
Inverse Coulomb permittivity $1/\epsi_C$ versus $K_0^2 \equiv k_0^2/2 eB$ for different Landau level regimes: $n_i = 0$--$3$ (a), $n_i = 7$--$10$ (b), and $n_i = 43$--$46$ (c). 
The black and blue lines represent the results at $q_s = 2.02 k_F$ and $\sqrt{eB}$, respectively. 
The green and red lines indicate the results at $2\sqrt{eB/3}$ and $\sqrt{eB/3}$ in the upper panel (a); $2\sqrt{eB/17}$ and $\sqrt{eB/17}$ in the middle panel (b); and $2\sqrt{eB/89}$ and $\sqrt{eB/89}$ in the lower panel (c), respectively. 
The solid, long-dashed, and dashed lines denote the results for $\theta_q = 0$, $\pi/4$, and $\pi/2$, respectively. The dotted lines show the corresponding results for $B=0$.
}
\label{PmB12}
\end{figure}

It is well known that quantum oscillations \cite{LK56} arise in strong magnetic fields 
as a consequence of Landau quantization, as exemplified by the de Haas–van Alphen \cite{dHvA1930} and Shubnikov–de Haas effects \cite{SHeff30}. 
The oscillatory behavior originates from the periodic modulation of the density of states 
as successive Landau levels cross the Fermi surface. 
In this work, we investigate how such oscillatory behaviors manifest themselves in the permittivity.

In Fig.~\ref{PmB12}, we show the inverse Coulomb permittivity $1/\epsi_C$ 
as a function of $K_0^2 \equiv k_0^2 / 2eB = n_i + K_F^2$ with $K_F^2 = k_F^2 / 2eB$. 
Here, $k_0$ represents the Fermi momentum along the magnetic field in the lowest Landau level (LLL, $n_i = 0$), which is expressed as
\begin{equation}
k_0^2 \equiv k_z^2 (0) = 2 eB n_i + k_F^2.
\end{equation}

The black lines represent the results for $q_s = 1.01 \times 2 k_F$, chosen to probe the vicinity of the Kohn anomaly. 
The blue, red, and green lines correspond to the results at $q_s = \sqrt{eB}$, $q_s = 2 \sqrt{eB / (2 n_L + 1)}$, 
and $q_s = \sqrt{eB / (2 n_L + 1)}$, respectively, 
where $n_L = 1$, $8$, and $44$ for the upper (a), middle (b), and lower (c) panels. 

The physical motivation for choosing these specific values of $q_s$ is as follows. 
In a strong magnetic field, the electron momentum in the $xy$-plane is discretized due to Landau quantization. Consequently, the characteristic momentum scales of the system are governed 
by the momentum difference associated with the magnetic length, $\sqrt{eB}$, 
and the Fermi momentum along the $z$-direction, $k_F$. 
From a semi-classical perspective, the characteristic spatial scale is given 
by the Larmor radius, $r_L \sim \sqrt{(2n_L+1) / eB}$, 
where $n_L$ is the electron Landau index. 
This implies that the characteristic momentum scale scales as $q_s \sim \sqrt{eB / (2n_L+1)}$, 
which justifies our choice for the parameters of the blue, red, and green curves.

As expected, Fig.~\ref{PmB12} confirms that $1/\epsi_C$ undergoes clear periodic quantum oscillations. 
As discussed previously, the one-loop RPA correlation function $G^C(0; \vq)$ 
diverges at $q_s = 2 k_F / \cos \theta_q$. 
For the parallel propagation case ($\theta_q = 0$), $1/\epsi_C$ drops sharply to approximately zero 
near the Kohn anomaly, $q_s \approx 2 \times 1.01 k_F$,  
which occurs when $K_0^2 = k_0^2 / 2eB$ takes an integer value.

Furthermore, within each single periodic region, $1/\epsi_C$ monotonically increases as $K_0^2$ (and hence the electron density) increases. This behavior stands in sharp contrast to the zero-field case ($B = 0$, represented by the dotted lines), where the inverse permittivity monotonically decreases with increasing momentum. 
Additionally, as the electron density increases further, the discrepancy between the finite-field and zero-field results systematically diminishes, demonstrating that the relative impact of the magnetic field becomes less prominent in the high-density regime.

%
%
\begin{figure}[hbt]
\centering
\includegraphics[scale=0.5,angle=270]{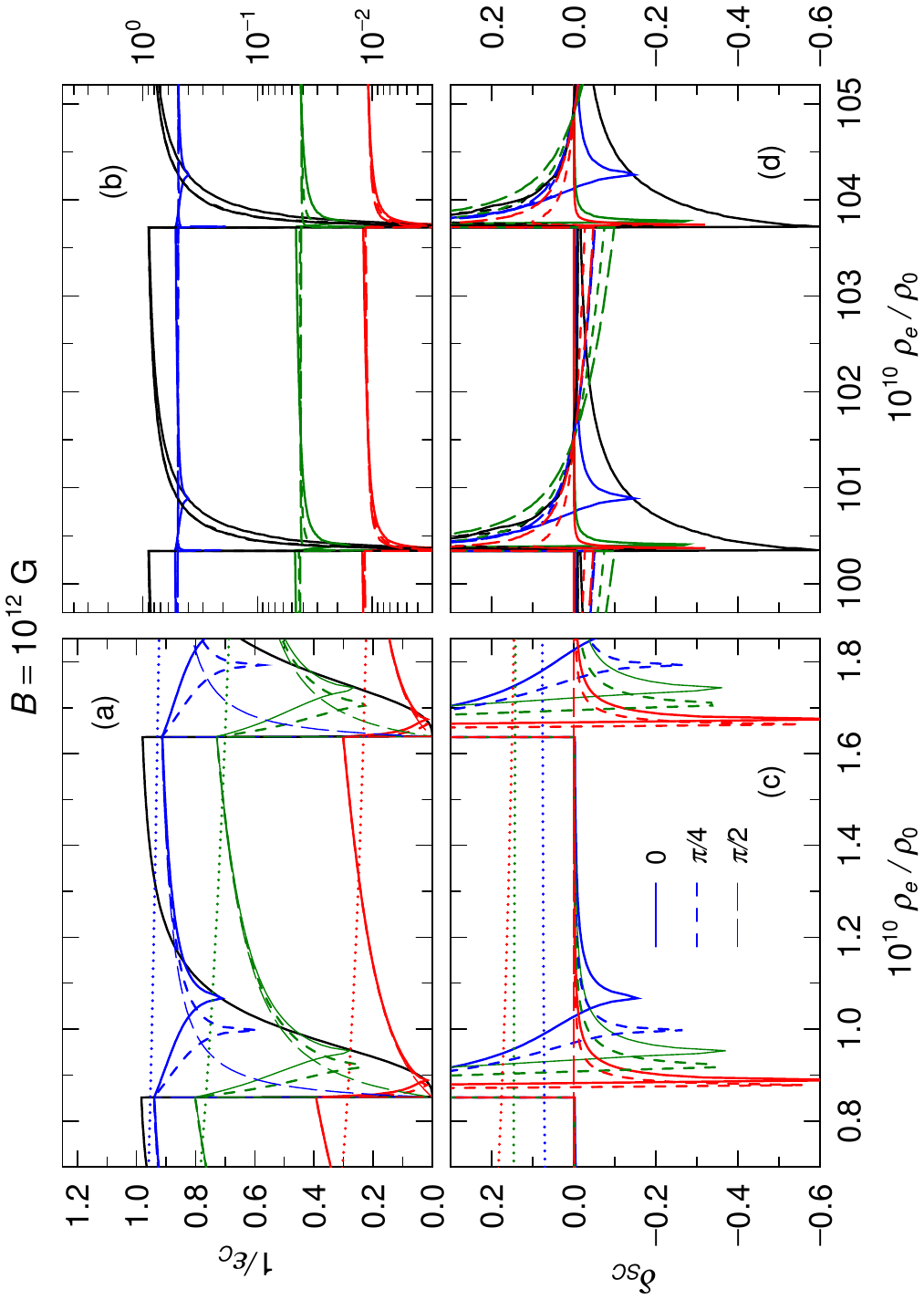}
\caption{\small
Electron density dependence of the inverse Coulomb permittivity $1/\epsi_C$ (top panels: a, b) 
and the screening deviation $\delta_{SC}$ (bottom panels: c, d) at $B = 10^{12}$~G. 
The left panels (a, c) represent the low-density region around $n_i = 2$, 
and the right panels (b, d) represent the high-density region around $n_i = 44$. 
The solid, dashed, and long-dashed lines denote the results for propagation angles 
of $\theta_q = 0$, $\pi/4$, and $\pi/2$, respectively. 
The black, blue, green, and red lines correspond to the specific values of $q_s$ defined in Fig.~\ref{PmB12}, 
and the dotted lines indicate the zero-field results ($B=0$).
}
\label{PmB12d}
\end{figure}

In order to examine these quantum effects in more detail, 
we show the direct density dependence of the inverse Coulomb permittivity $1/\epsi_C$ (a, b) 
and the screening deviation $\delta_{SC}$ (c, d) in Fig.~\ref{PmB12d}. 
Panels (a) and (c) focus on the low-density region around the threshold $n_i = 2$, 
while panels (b) and (d) display the high-density region around $n_i = 44$, 
under a strong magnetic field of $B = 10^{12}$~G. 
These threshold structures are loosely analogous to shell effects in finite nuclei.

Here, we notice again that both $1/\epsi_C$ and $\delta_{SC}$ exhibit distinct discontinuities at the exact density thresholds where $k_F = 0$. 
In the immediate vicinity of these discontinuous points, the inverse permittivities drastically increase, 
approaching unity, while $\delta_{SC}$ drops sharply toward negative values. 
This characteristic behavior stems from the fact that the Debye screening mass $M_D$ diverges at $k_F = 0$, 
where the density of states at the Fermi surface becomes singular 
due to the one-dimensional nature of the Landau levels. 
It should be noted, however, that this localized divergence is a mathematical consequence 
of the ideal unmitigated state density and does not 
alter the fundamental collective screening features away from the threshold.

Furthermore, Fig.~\ref{PmB12d} reveals a rich angular ($\theta_q$) and momentum ($q_s$) dependence. 
In the low-density regime [panels (a) and (c)], the anisotropy introduced by the magnetic field is highly pronounced. 
For the parallel propagation case ($\theta_q = 0$, solid lines), the quantum oscillations achieve their maximum amplitude, demonstrating that the screening is highly sensitive to the longitudinal Fermi motion. 
As the angle $\theta_q$ increases toward the perpendicular direction ($\pi/2$, long-dashed lines), 
the amplitude of these oscillations systematically suppresses, 
and the curves converge toward the unmagnetized $B=0$ results (dotted lines), 
where the spatial anisotropy is washed out.


\begin{figure}[hbt]
\centering
\includegraphics[scale=0.6]{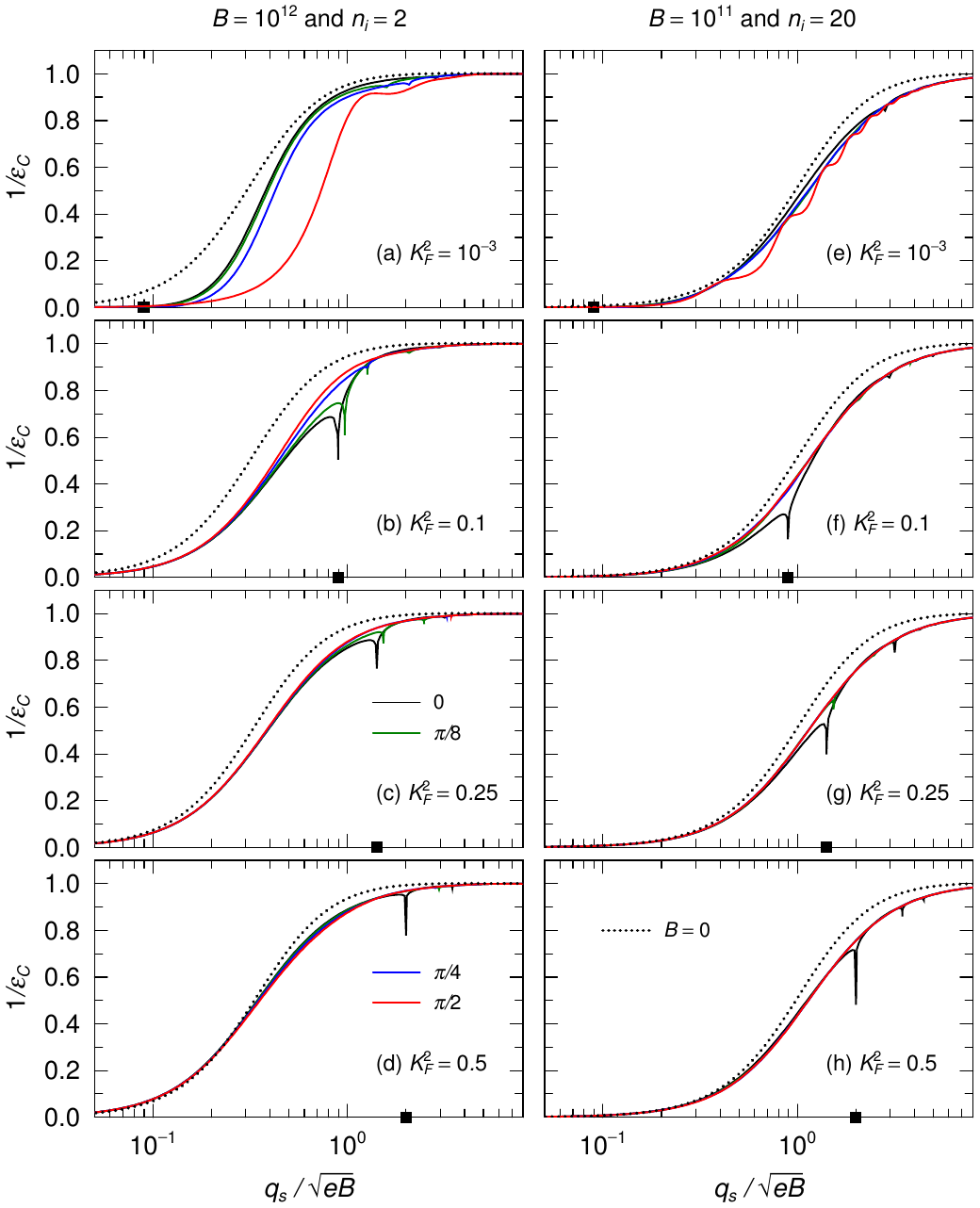}
\caption{\small
Inverse Coulomb permittivity $1/\epsilon_C$ versus normalized momentum transfer $q_s/\sqrt{eB}$ 
for $\rho_e/\rho_0 \approx 10^{-10}$. Left panels (a)–(d) show $B = 10^{12}$~G ($n_F = 2$) 
and right panels (e)–(h) show $B = 10^{11}$~G ($n_F = 20$), 
for various Fermi momentum parameters $K_F^2$. 
Solid lines indicate polar angles $\theta_q = 0, \pi/8, \pi/4,$ and $\pi/2$. 
Dotted lines show the $B = 0$ case. 
Squares on the horizontal axes mark $q_s = 2k_F$, where the 1D Kohn anomaly forms a sharp cusp.
}
\label{PmRM10}
\end{figure}

In the higher-density regime [panels (b) and (d)], a clear qualitative shift is observed. Although the periodic quantum oscillations still persist at the threshold points, the overall modulation amplitude is significantly reduced compared to the $n_i = 2$ case. In this regime, the higher Fermi energy dominates over the magnetic quantization energy scale ($E_F \gg \hbar \omega_c$). Consequently, the dielectric responses for all angles and momenta smoothly merge into a single tight bundle that closely tracks the zero-field baseline, indicating that the relative impact of the strong magnetic field becomes systematically obscured in ultra-dense degenerate electron gases.

In Fig.~\ref{PmRM10}, we show the momentum transfer dependence of the inverse Coulomb permittivity $1/\epsi_C$ around the electron density region $\rho_e / \rho_0 \approx 10^{-10}$ under strong magnetic fields of $B=10^{12}$~G (left panels) and $B=10^{11}$~G (right panels). 
Overall, the inverse permittivities under strong magnetic fields are found to be systematically smaller than those in the field-free case ($B=0$). 
This suppression indicates that the electronic screening effect is enhanced relative to the $B=0$ case  by the Landau quantization. 
Particularly at $B=10^{12}$~G and $K_F^2 = 10^{-3}$  (a), 
$1/\epsi_C$ is highly suppressed in the low-$q_s$ region across all polar angles, 
with the suppression being most pronounced at $\theta_q = \pi/2$ for $q_s \lesssim \sqrt{eB}$.

Furthermore, when $\theta_q = 0$, distinct cusps appear exactly at $q_s = 2 k_F$, 
as indicated by the full squares. 
Mathematically, $1/\epsi_C$ is expected to vanish at this characteristic momentum transfer 
due to the Kohn anomaly. 
Although the finite resolution in our numerical discretization prevents $1/\epsi_C$ from dropping precisely to zero, the numerical results clearly capture the logarithmic derivative singularity characteristic of the anomaly. 

In the small-$q_s$ region, the long-wavelength Debye response dominates, 
resulting in $1/\epsilon_C \ll$ 1, and the finite-q Kohn anomaly is masked.

%
\begin{figure}[hbt]
\centering
\centering
\includegraphics[scale=0.5,angle=270]{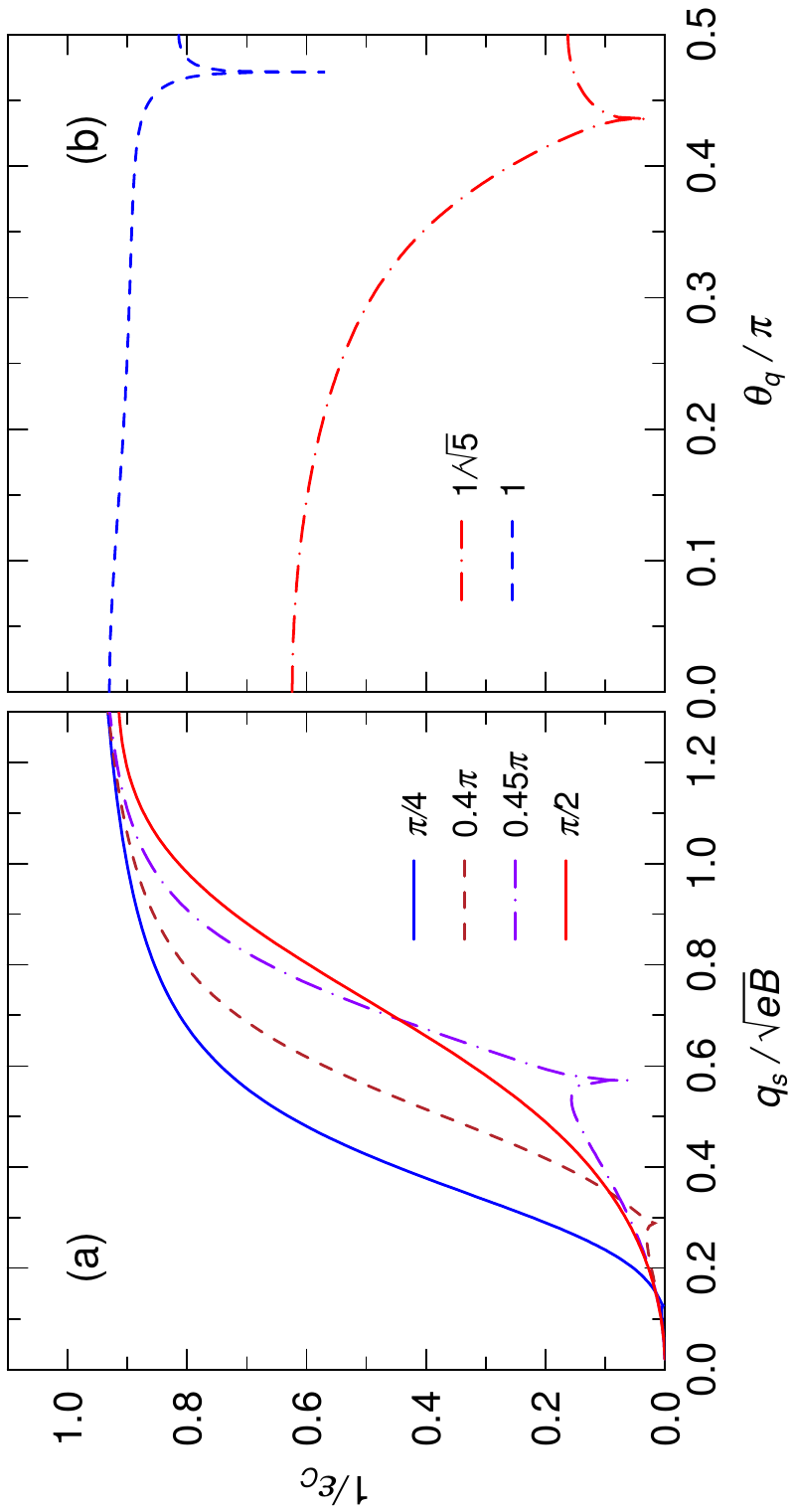}
\caption{\small
Momentum and angular dependence of the inverse Coulomb permittivity $1/\epsilon_C$ at $B = 10^{12}$~G for $n_i = 2$ and $K_F^2 = 0.001$. (a) $1/\epsilon_C$ as a function of the normalized momentum transfer $q_s/\sqrt{eB}$ at various fixed angles $\theta_q$. The blue solid, brown dashed, purple dot-dashed, and red solid lines correspond to $\theta_q = \pi/4$, $0.4\pi$, $0.45\pi$, and $\pi/2$, respectively. (b) $\theta_q$ dependence at fixed momentum transfers. 
The red dash-dotted and blue dashed lines represent 
the results for $q_s/\sqrt{eB} = 1/\sqrt{5}$ and $1$, respectively.
}
\label{PmB12K0}
\end{figure}

To evaluate the anisotropy of the Coulomb permittivity in momentum space, the explicit momentum-magnitude and angular dependencies for the low-density case of $K_F^2 = 0.001$ at $B=10^{12}$~G are illustrated in Fig.~\ref{PmB12K0}. 
While our previous comprehensive results showed a distinct and significantly different behavior exclusively at the perpendicular direction ($\theta_q = \pi/2$), the current analysis aims to elucidate the continuous transition in the intermediate angular region between $\theta_q = \pi/4$ and $\pi/2$.

As shown in the left panel (a), the momentum-magnitude dependence reveals that the screening effect becomes progressively enhanced (i.e., $1/\epsilon_C$ decreases) as the angle approaches to $\pi/2$. This behavior is non-trivial, since the conventional $LLL$ picture emphasizes the strong longitudinal response along the magnetic field. 
The present result indicates that finite-momentum screening and 
inter-Landau-level contributions can modify this simple intuition.
Notably, characteristic cusp structures emerge in this momentum dependence at specific angles: at $q_s/\sqrt{eB} = 0.2809$ for $\theta_q = 0.4\pi$, and at $q_s/\sqrt{eB} = 0.5176$ for $\theta_q = 0.45\pi$. 
This transition and the angular evolution of these features are further confirmed in the right panel (b) via the explicit $\theta_q$ dependence at fixed momentum transfers. 
In the forward scattering limit ($\theta_q \to 0$), the curves systematically approach the behavior governed by the strong dimensional reduction and the resulting Kohn anomaly induced by the magnetic field, providing a comprehensive picture of how the interplay between momentum magnitude and orientation dictates the anisotropic screening behavior.

It should be noted that for a higher density region within the single Landau level where $K_F^2 \gtrsim 0.5$, the Kohn anomaly momentum $q_s = 2 k_F$ shifts into the high-momentum saturation regime where $1/\epsi_C \approx 1$. 
Near this fully occupied limit of the current Landau level, the dynamic screening features and the anisotropy are largely suppressed, converging toward the trivial free-Coulomb behavior. 
Therefore, to clearly elucidate the prominent anisotropic screening effects and the characteristic quantum singular behavior, such as the perfect screening ($1/\epsi_C = 0$) satisfying the one-dimensional nesting condition $q_s = 2 k_F/\cos \theta_q$, we focus our discussion on the lower carrier density regime in this work.

\begin{figure}[hbt]
\centering
\includegraphics[scale=0.6]{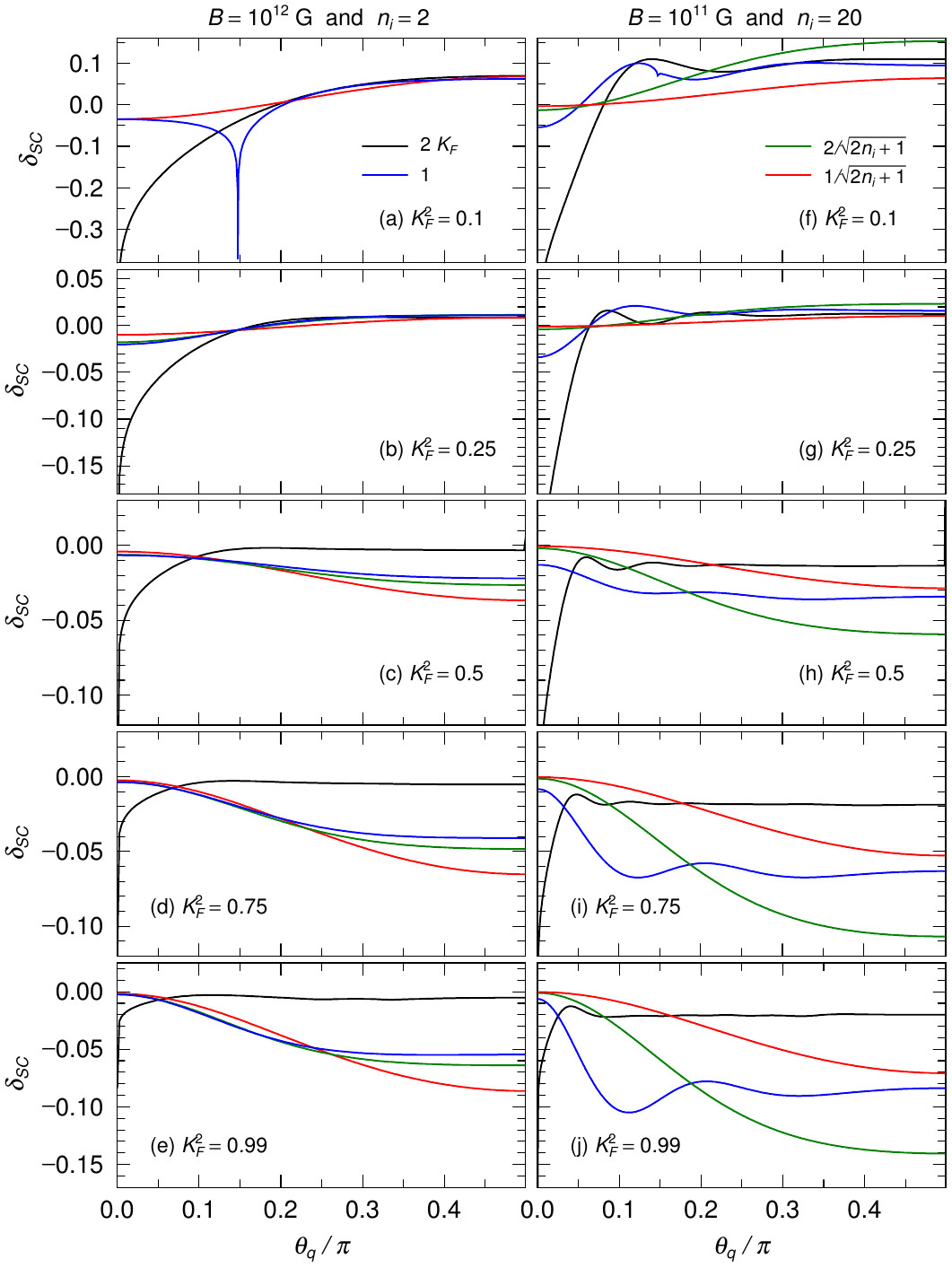}
\caption{\small 
Polar angle dependence of the screening deviation $\delta_{SC}$ around the electron density region $\rho_e / \rho_0 \approx 10^{-10}$. 
The left panels (a--e) show the results for $n_i = 2$ at $B = 10^{12}$~G, while the right panels (f--j) show those for $n_i = 20$ at $B = 10^{11}$~G. From top to bottom, the rows correspond to $K_F^2 = 0.1$ (a, f), $0.25$ (b, g), $0.5$ (c, h), $0.75$ (d, i), and $0.99$ (e, j), respectively. The black, blue, green and red lines represent the results at fixed momentum transfers of $q_s = 2 k_F$, $\sqrt{eB}$, $2 \sqrt{eB/(2n_i + 1)}$, and $\sqrt{eB/(2n_i + 1)}$, respectively.
}
\label{PmAg}
\end{figure}

In order to investigate the anisotropy of the electronic screening, in Fig.~\ref{PmAg} we show the polar angle dependence of the screening deviation $\delta_{\text{SC}}$ 
around the electron density region $\rho_e / \rho_0 \approx 10^{-10}$ for $B = 10^{12}$~G ($n_F = 2$, left panels) and $B = 10^{11}$~G ($n_F = 20$, right panels). 
When the carrier density is relatively low ($K_F^2 \le 0.25$ within the currently occupied Landau level), the results exhibit an almost isotropic behavior with very small angular dependence, except in the immediate vicinity 
of the forward scattering limit ($\theta_q = 0$)
 where the Kohn anomaly dictates the singular behavior. 
On the other hand, as the density increases toward the completion of the Landau level occupation ($K_F^2 \ge 0.5$), a pronounced anisotropic behavior emerges, characterized by a systematic decrease in $\delta_{\text{SC}}$ with increasing $\theta_q$. 

This distinct anisotropy becomes practically decoupled from the prominent features of the Kohn anomaly in this higher density regime. 
Mathematically, the Kohn anomaly condition $q_z = 2k_F$ corresponds to the normalized momentum transfer of $q_s / \sqrt{eB} = 2\sqrt{2}K_F / \cos\theta_q$. 
As the carrier density increases beyond $K_F^2 > 0.5$, the characteristic momentum scale for the Kohn anomaly shifts to higher values, moving into the asymptotic momentum regime where the overall screening effect is already substantially diminished ($\delta_{\text{SC}} \to 0$). 
Consequently, rather than dominating the overall angular profile, the singular contribution of the Kohn anomaly becomes less prominent over the dominant screening momentum range, leaving the smooth, orientation-dependent spatial anisotropy to govern the system.

\begin{figure}[hbt]
\centering
\includegraphics[scale=0.6]{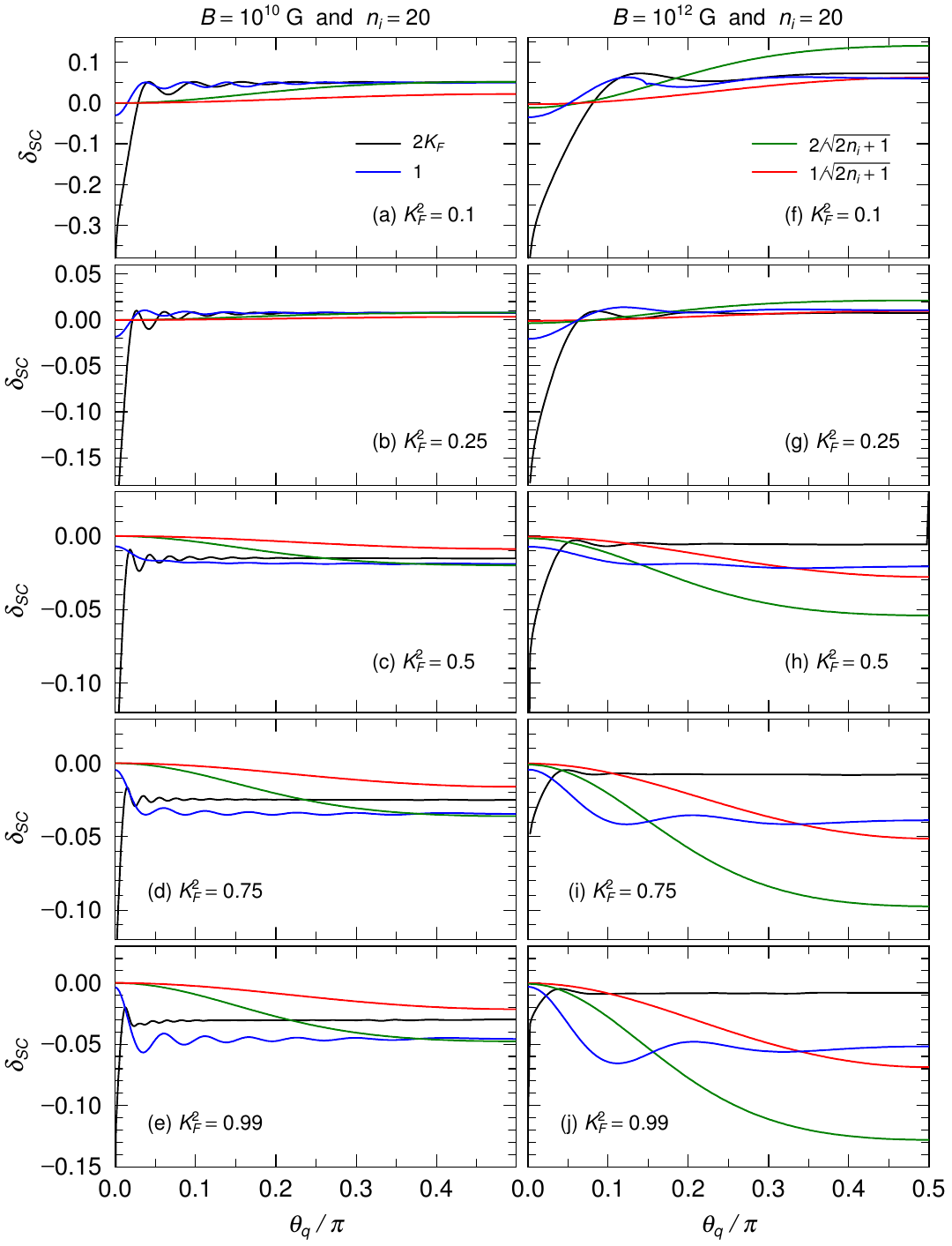}
\caption{\small
Polar angle dependence of the screening deviation $\delta_{SC}$ 
with the maximum Landau level index fixed at $n_i = 20$. 
The left panels (a--e) represent the results for $B = 10^{10}$~G, and the right panels (f--j) represent those for $B = 10^{12}$~G.
The density rows for $K_F^2$ from top to bottom are identical to those in Fig.~\ref{PmAg}. The line specifications are the same as in Fig.~\ref{PmAg}.
}
\label{PmAgL20}
\end{figure}

Furthermore, comparing the two magnetic field strengths in Fig.~\ref{PmAg}, 
the weaker magnetic field ($B = 10^{11}$~G) exhibits more intricate structures and 
a more pronounced anisotropy than the stronger field ($B = 10^{12}$~G). 
This behavior is attributed to the larger number of occupied Landau levels participating 
in the screening processes at weaker fields, 
which introduces additional quantum scales into the momentum dependence \cite{GC1968}.

To systematically isolate the role of the Landau-level index from the pure density variation, we illustrate in Fig.~\ref{PmAgL20} the results for a fixed maximum occupied i
ndex $n_F = 20$ at $B = 10^{12}$~G ($\rho_e / \rho_0 \approx 3.0\text{--}3.2 \times 10^{-9}$) 
and $B = 10^{10}$~G ($\rho_e / \rho_0 \approx 3.0\text{--}3.2 \times 10^{-12}$).
It should be noted that this comparison does not represent a pure magnetic field variation 
at constant density; rather, fixing $n_F$ implies that the electron density scales down 
with $B$, accompanied by changes in the dimensionless ratio $m_e^2/eB$, 
individual Fermi momenta $k_z(n)$, and the transverse momentum-transfer form factors.
While the quantitative differences between $B = 10^{11}$~G (the right panels in Fig.~\ref{PmAg}) 
and $B = 10^{12}$~G remain relatively minor, a qualitative evolution is observed 
as the field drops further to $B = 10^{10}$~G. 
At this lower field, although the global angular trend remains qualitatively consistent, 
the negative features associated with the Kohn anomaly near $\theta_q = 0$ 
become significantly narrower, 
and the overall anisotropy in momentum space is visibly diminished.

Furthermore, a high-frequency oscillatory behavior appears 
at small angles ($\theta_q/\pi \lesssim 0.15$) for $B = 10^{10}$~G.
This structure reflects the quantum oscillations induced by the simultaneous participation of multiple occupied Landau levels ($n \le n_F$), 
where each level possesses its own distinct Fermi momentum $k_z(n)$ that satisfies the one-dimensional nesting condition at different angular positions \cite{Horing69}. 
Concurrently, in the large-angle region ($\theta_q \to \pi/2$), the curves for different momentum transfers tend to converge toward a nearly uniform value, signifying that the system gradually recovers its isotropic, three-dimensional character as the magnetic field weakens \cite{Potekhin99}.

The quantum oscillatory behavior of the screening properties originates from the quantum oscillations of the electronic compressibility. 
Since the Debye screening mass is proportional to the compressibility, 
$M_D^2 = e^2 (\partial n_e / \partial \mu)$, and $\partial n_e / \partial \mu = g(E_F)$ is directly proportional to the density of states at the Fermi surface, 
each Landau-level occupation threshold produces 
a singular enhancement of the density of states at $E_F$. 
Consequently, the compressibility, Debye mass, and 
static dielectric response exhibit identical quantum oscillatory structures.

Since the present analysis is restricted to the static density response $G_{00}(0, \bm{q})$, 
omitting the QED vacuum polarization affects only the high-momentum vacuum contribution 
and does not alter the medium-induced screening and compressibility properties evaluated here. 


\section{Summary}

The electronic screening behavior under strong magnetic fields exhibits several distinct features governed by the quantum periodic nature of Landau levels and spatial anisotropy. 
First, owing to the inherent periodicity in Landau-level occupation, the screening properties are found to be relatively insensitive to variations in the carrier density compared to those induced by changes in the magnetic field strength. 
Within a single occupation cycle, the Kohn anomaly plays a dominant role in the lower-density regime ($K_F^2 \lesssim 0.25$), leading to a severe suppression of the inverse permittivity; specifically, $\delta_{\text{SC}}$ approaches $-1$. 
However, in the immediate vicinity just above the boundary ($K_F^2 \to 0^+$), a unique and pronounced angular effect emerges. 

Furthermore, a subtle competition in the orientation-dependent screening is observed: 
while an enhanced screening effect appears perpendicular to the magnetic field in the extreme limits of $K_F^2 \ll 1$ and $K_F^2 \gtrsim 0.25$, the screening is instead enhanced parallel to the field direction in the intermediate regime. 
It is worth emphasizing that while the quantum singular effect driven by the Kohn anomaly is dramatic and dominant at lower densities, the latter orientation-dependent enhancements remain a subtle modulation on the order of a few percent. 
These findings establish a comprehensive and quantitative understanding of field-induced screening anisotropy, demonstrating that the screening response in a magnetized electron gas is controlled not only by the carrier density but also by the specific phase within the Landau-level occupation cycle.

The anisotropic and momentum-dependent static screening properties identified in this work 
have direct implications for physical processes in strongly magnetized environments, 
such as electron-impurity scattering, transport coefficients, phonon spectrum modulations, 
and lattice stability in magnetar crusts \cite{Potekhin99,Chamel2008,Baiko2009}. 
In particular, the direction-dependent screening potential modifies the effective ion-ion interactions 
and pycnonuclear reaction rates in dense stellar interiors \cite{Horing69}.

While the present formulation is restricted to the static limit ($q_0 = 0$), a natural and crucial extension of this framework is the inclusion of dynamical screening effects ($q_0 \neq 0$). 
In high-density plasmas and intense magnetic fields, collective modes such as magnetoplasmons, Bernstein modes, and directional Landau damping play vital roles in energy transport and non-equilibrium scattering processes \cite{Horing69,Sharma2010,Fukushima2025}. 
Evaluating the full frequency- and momentum-dependent dielectric function $\epsilon(q_0, \bm{q})$ incorporating arbitrary Landau-level transitions, as well as the coupled electromagnetic mode structure beyond the uncoupled Coulomb channel, remains an essential direction for future investigation.

\acknowledgements
TM is supported in part by Grants-in-Aid for Scientific Research of JSPS (24K07057). 
ABB is supported in part by the U.S. National Science Foundation grant No. PHY-2411495.    
M.-K.C. was also supported by the NRF Basic Science Research Program under
Grant Nos. RS-2021-NR060129 and RS-2025-16071941.
TK is supported in part by the National Key R\&D Program of China (2022YFA1602401) and the National Natural Science Foundation of China (No. 12335009 \& 12435010).

\section*{Data Availability}

No data were created or analyzed in this study.

\end{document}